\documentclass[aps,showpacs,nofootinbib,superscriptaddress]{revtex4}

\usepackage{graphicx}
\usepackage{amsmath}
\usepackage{amssymb}
\def\slashchar#1{\setbox0=\hbox{$#1$}
   \dimen0=\wd0 \setbox1=\hbox{/} \dimen1=\wd1
   \ifdim\dimen0>\dimen1 \rlap{\hbox to \dimen0{\hfil/\hfil}} #1
   \else  \rlap{\hbox to \dimen1{\hfil$#1$\hfil}} / \fi}
\begin{document}
\title{$\nu$ induced threshold production of two pions and $N^*(1440)$ 
  electroweak form factors}

\author{E. Hern\'andez} 
\affiliation{Grupo de F\'\i sica Nuclear,
Departamento de F\'\i sica Fundamental e IUFFyM,\\ Universidad de
Salamanca, E-37008 Salamanca, Spain.}  
\author{J. Nieves}
\affiliation{Departamento de F\'\i sica At\'omica, Molecular y Nuclear,
\\Universidad de Granada, E-18071 Granada, Spain.}
\author{S.K. Singh} 
\affiliation{Department of Physics, Aligarh Muslim
University, Aligarh-202002, India.}  
\author{M. Valverde}
\affiliation{Departamento de F\'\i sica At\'omica, Molecular y Nuclear,
\\Universidad de Granada, E-18071 Granada, Spain.}
\author{M.J. \surname{Vicente Vacas}} 
\affiliation{Departamento de
F\'\i sica Te\'orica and IFIC, Centro Mixto Universidad de
Valencia-CSIC\\ Institutos de Investigaci\'on de Paterna,
Aptdo. 22085, E-46071 Valencia, Spain.}

\today
\begin{abstract}
We study the threshold production of two pions induced by neutrinos in
nucleon targets. The contribution of nucleon, pion and contact terms
are calculated using a chiral Lagrangian. The contribution of the
Roper resonance, neglected in earlier studies, has also been taken
into account. The numerical results for the cross sections are
presented and compared with the available experimental data. It has
been found that in the two pion channels with $\pi^+\pi^-$ and
$\pi^0\pi^0$ in the final state, the contribution of the $N^*(1440)$
is quite important and could be used to determine the $N^*(1440)$
electroweak transition form factors if experimental data with better
statistics become available in the future.

\end{abstract}


\maketitle

%=========================================

\section{Introduction}

The important ongoing experimental effort addressing the questions of
neutrino oscillations is bringing out, as a fortunate byproduct, much
information on the structure of hadrons and nuclei. Apart from the
intrinsic interest of the knowledge of axial form factors, structure
functions or the strange quark content of the nucleon, a proper and
precise understanding of various processes induced by neutrino
interactions is required in the experimental analysis of background
substraction, $\nu$-flux determination and particle identification in
the neutrino oscillation experiments.

In the study of the neutrino--nucleus interaction, we can distinguish
several energy regions. At low energies we have quasielastic
scattering (QE) in which a nucleon is knocked out of the target
nucleus. There has been an strong effort both experimental and
theoretical to measure and describe this process
\cite{Nieves:2004wx,Nieves:2005rq,Valverde:2006zn,Martinez:2005xe,Meucci:2004ip,Caballero:2005sj,Leitner:2006ww,Gran:2006jn,:2007ru}. The
basic ingredient is the relatively well known neutrino nucleon elastic
interaction, although nuclear effects like Fermi motion, Pauli
blocking or long range RPA correlations are also needed. These nuclear
effects have been found to strongly modify the cross section and also
to distort angular and energy distributions of the final particles.

At intermediate energies, above 0.5 GeV, one pion production becomes
relevant. The knowledge of the elementary process $\nu +N\to
\ell+N'+\pi$ is not so well established. One of the reasons is the
scarcity of data \cite{Radecky:1981fn,Kitagaki:1990vs}. Most of the
theoretical models assume the dominance of $\Delta(1232)$ resonance
mechanisms
\cite{Adler:1968tw,Schreiner:1973ka,AlvarezRuso:1997jr,AlvarezRuso:1998hi,Paschos:2003qr,Lalakulich:2005cs,Lalakulich:2006sw}
but others also include background
terms\cite{Fogli:1979cz,Sato:2003rq,Hernandez:2007qq}.  The major
uncertainties of these models appear in the $N\Delta(1232)$ transition
axial form factors that are fitted, with some theoretical ansatz to
the available data.  It is hoped that new data on single pion
production from neutrino experiments at K2K \cite{nuint07} and
MiniBooNE \cite{Wascko:2006tx,link} could help to determine these form
factors.

In nuclei, the description of pion production requires a realistic
model to account for the final state interaction (FSI) of the
pion. This is usually implemented in MonteCarlo codes. However, that
is not enough. It has been shown in several works that also the
production mechanisms are modified in the medium. Furthermore, in some
particular cases, like the coherent pion production, a quantum
treatment of FSI is necessary. In any case, there is clear
progress in our understanding of this reaction and some recent
\cite{Hasegawa:2005td} and coming data \cite{Wascko:2006ty} may
already put strong constraints on the $\Delta$ form factors
\cite{AlvarezRuso:2007it}.

Above these energies, but still below the DIS region, new inelastic
channels are open and several baryonic resonances beyond the
$\Delta(1232)$ can be excited
\cite{Rein:1980wg,Lalakulich:2006sw}. Recently, there has been an
important progress in the determination of their vector form factors
with the advent of high quality electromagnetic data
\cite{Burkert:2004sk,Aznauryan:2004jd,Tiator:2003uu}. Our knowledge of
the axial form factors is, in general, poorer due to the scarcity of
experimental information \cite{Lalakulich:2006sw}.

The first of these resonances is the $N^*(1440)$, for which the weak
excitation can allow to study the axial sector.  Although its main
decay channel is to $N\pi$, its contribution to the one pion
production cross section has been found to be negligible
\cite{AlvarezRuso:1997jr} because of the strong dominance of $\Delta$
mechanisms. However, the situation can be different for the production
of two pions.  This channel starts at invariant masses of the hadronic
sector just below the $\Delta$. However, the $\Delta$ does not couple
to two pions in $s$-wave and thus it is not very relevant at these
energies, where only slow pions are produced. On the other hand, the
Roper resonance $N^*(1440)$ has a sizable decay into a scalar pion
pair and it is very wide so that its contribution could be
large. Indeed, Roper excitation mechanisms are known to play a major
role in other two pion production reactions close to threshold, like
$\pi N\to\pi\pi N$ \cite{Oset:1985wt,Kamano:2004es,Hernandez:2002xk}
or $N N\to\pi\pi N N$ \cite{Alvarez-Ruso:1997mx,Alvarez-Ruso:1998xg}.

In the weak interaction sector, there exist very few attempts to study
the two pion production induced by neutrinos and antineutrinos. The
older experiments done at CERN~\cite{Allasia:1990uy,Jones:1991tm,Wittek:1989tu,Grassler:1985cd}, in the regime of high energy have
studied two and three pion production to investigate the diffractive
production of meson resonances like $\rho$ and $\eta$. The later
experiments done at ANL \cite{Barish:1978pj,Day:1984nf} and BNL
\cite{Kitagaki:1986ct} at lower energies have investigated the two
pion production processes, specially in the threshold region, in order
to test the predictions of the chiral symmetry of the strong
interaction Lagrangian. Such studies were theoretically proposed by
Biswas et al.~\cite{Biswas:1978ey} and Adjei et
al.~\cite{Adjei:1980nj,Adjei:1981nw}.  Biswas et al.\ used PCAC and
current algebra methods to calculate the threshold production of two
pions. On the other hand, the work of Adjei et al.\ made specific
predictions for the threshold production of two pions in a restricted
kinematic region using an effective Lagrangian incorporating chiral
symmetry and its breaking, governed by a free parameter ($\xi$).
Imposing these kinematical restrictions, the experimental data of ANL
and BNL were analyzed and compared with Adjei et al.\
results. However, the model did not include any resonance production
and the contribution of the $N^*(1440)$ should be taken into account.
Furthermore we use an expansion of the chiral Lagrangian that includes
terms up to ${\cal O}(1/f_\pi^3)$, while Adjei et al.\ kept only terms
up to ${\cal O}(1/f_\pi^2)$.  Now that the pattern of chiral symmetry
breaking is well known and the background terms contribution to the
threshold production of two pions is fully determined, the process
could be used to study the electroweak transition form factors of the Roper
resonance.

In this paper, we will study the $\nu_l N \to l^- \pi\pi N$
channel close to threshold.  Apart from Roper resonance contribution,
many other background mechanisms, that only involve nucleons and
pions, appear and are described by an effective Lagrangian. We will
use the lowest order chiral perturbation theory Lagrangian to derive
the needed axial and vector currents.

In Sec.~\ref{sec:CC}, we present the formalism and the Lagrangians
used in our model. We also give the expressions of the Roper form
factors. In Sec.~\ref{sec:res}, we present our results and compare them
with the available data. Finally, the appendix gives the detailed
formulas of the contributions of the background mechanisms for all
channels.

\section{Model for $\nu$ induced two pion production}
\label{sec:CC}

\subsection{Kinematics}
\label{sec:kin}

We will focus on the neutrino--pion production reaction off the
nucleon driven by charged currents,
\begin{equation}
\nu_l(k) + N(p) \to l^-(k^\prime) + N(p^\prime) + \pi(k_{\pi_1})+\pi(k_{\pi_2})
\label{eq:reac}
\end{equation}
though the generalization of the obtained expressions to antineutrino
induced reactions is straightforward.
The unpolarized differential cross section, with respect to the
outgoing lepton kinematical variables, is given in the
Laboratory (LAB) frame by
\begin{equation}
\frac{d\sigma_{\nu_ll}}{d\Omega(\hat{k^\prime})dE^\prime} = 
 \frac{G^2}{4\pi^2}\frac{|\vec{k}^\prime|}{|\vec{k}|}
  L_{\mu\sigma}\left(W^{\mu\sigma}_{{\rm CC}2\pi}\right)
\label{eq:sec}
\end{equation}
with $\vec{k}$ and $\vec{k}^\prime$ the LAB lepton momenta,
$E^{\prime} = (\vec{k}^{\prime\, 2} + m_l^2 )^{1/2}$ and $m_l$ the
energy and the mass of the outgoing lepton, 
%($m_\mu = 105.65$ MeV, $m_e = 0.511$ MeV )
$G=1.1664\times 10^{-11}$ MeV$^{-2}$, the Fermi
constant. We take $\epsilon_{0123}= +1$ and the metric $g^{\mu\nu}=(+,-,-,-)$, thus
the leptonic tensor is given by:
\begin{equation}
L_{\mu\sigma} = (L_s)_{\mu\sigma} \pm i
 (L_a)_{\mu\sigma} =
 k^\prime_\mu k_\sigma +k^\prime_\sigma k_\mu
- g_{\mu\sigma} k\cdot k^\prime \pm 
i\epsilon_{\mu\sigma\alpha\beta}k^{\prime\alpha}k^\beta \label{eq:lep}
\end{equation}
%
%and it is not orthogonal to $q^\mu$ even for massless neutrinos, i.e,
%$L_{\mu\sigma}^{(\nu)} q^\mu = -m^2_l k_\sigma$.
where the $+(-)$ sign corresponds to neutrino(antineutrino) induced processes.

The hadronic tensor  reads as
\begin{multline}
W^{\mu\sigma}_{{\rm CC} 2\pi} = \\
 \overline{\sum_{\rm spins}} \int
\frac{d^3p^\prime}{(2\pi)^3} \frac{M}{E^\prime_N}
\frac{d^3k_{\pi_1}}{(2\pi)^3} \frac{1}{2E_{\pi_1}}
\frac{d^3k_{\pi_2}}{(2\pi)^3} \frac{1}{2E_{\pi_2}}
(2\pi)^3 \delta^4(p^\prime + k_{\pi_1} + k_{\pi_2}  - q - p) 
\langle N^\prime\pi_1\pi_2|j^\mu_{\rm cc+}(0)|N\rangle
\langle N^\prime\pi_1\pi_2|j^\sigma_{\rm cc+}(0)|N\rangle^*
\label{eq:wmunu}
\end{multline}
with $M$ the nucleon mass, 
$q=k-k^\prime$ and $E^\prime_N$ the energy of the
outgoing nucleon. The bar over the sum of initial and final spins,
denotes the average on the initial ones. As for the baryon
states, they are normalized so that 
$\langle \vec{p}\,|\vec{p}\,^\prime \rangle = (2\pi)^3 
\delta^3(\vec{p}-\vec{p}\,^\prime)p_0/m$.
By construction, the hadronic tensor can be split in
\begin{equation}
\left (W^{\mu\sigma}_{{\rm CC} 2\pi}\right) = 
\left (W^{\mu\sigma}_{{\rm CC} 2\pi}\right)_s + {\rm i}
\left(W^{\mu\sigma}_{{\rm CC} 2\pi}\right)_a 
\end{equation}
with $\left(W^{\mu\sigma}_{{\rm CC} 2\pi}\right)_s$
and $\left (W^{\mu\sigma}_{{\rm CC} 2\pi}\right)_a$ real
symmetric  and antisymmetric parts, respectively.

\subsection{Lagrangians for the non resonant terms}
\label{sec:lag1}

For the derivation of the hadronic tensor we use the effective
Lagrangian of the SU(2) non-linear $\sigma$ model. This model was used
previously in~\cite{Hernandez:2007qq} for the description of the
non-resonant contributions to one pion weak production processes off
nucleon. We refer the reader to that paper for details.  Up to ${\cal
O}(1/f_\pi^3)$, this SU(2) chiral Lagrangian reads
\begin{equation}
{\cal L} = \bar\Psi [{\rm i} \slashchar{\partial}-M] \Psi+\frac12
\partial_\mu\vec{\phi}\partial^\mu\vec{\phi} -\frac12 m_\pi^2
\vec{\phi}^{\,2} + {\cal L}_{\rm int}^\sigma \label{eq:lsigma}
\end{equation}
\begin{eqnarray}
{\cal L}_ {\rm int}^\sigma&=& \frac{g_A}{f_\pi} \bar\Psi \gamma^\mu\gamma_5
\frac{\vec{\tau}}{2}(\partial_\mu \vec{\phi})\Psi
-\frac{1}{4f_\pi^2}\bar\Psi \gamma_\mu \vec{\tau}\left
(\vec{\phi}\times\partial^\mu\vec{\phi}\right)\Psi -
\frac{1}{6f_\pi^2} \left ( \vec{\phi}^{\,2}
\partial_\mu\vec{\phi}\partial^\mu \vec{\phi}-(\vec{\phi}\partial_\mu
\vec{\phi} )(\vec{\phi}\partial^\mu
\vec{\phi}) \right) + \frac{m_\pi^2}{24f_\pi^2}(\vec{\phi}^{\,2})^2
\nonumber \\
&&-\frac{g_A}{6f_\pi^3} \bar\Psi \gamma^\mu\gamma_5
\left[\vec{\phi}^{\,2} \frac{\vec{\tau}}{2}\partial_\mu \vec{\phi} -
(\vec{\phi}\partial_\mu\vec{\phi})\frac{\vec{\tau}}{2} \vec{\phi}
\right]\Psi \, ,
\label{eq:lint}
\end{eqnarray}
where $\Psi= \left (\begin{array}{c}p\cr n\end{array}\right )$ is the
nucleon field, $\vec{\phi}$ is the isovector pion field, $\vec{\tau}$ are the Pauli matrices 
and $f_\pi=93$ MeV is the pion decay constant. 
The vector and axial currents generated from the Lagrangian in
Eq.~(\ref{eq:lsigma}) are given by~\cite{Hernandez:2007qq}
\begin{equation}
{\vec V}^\mu = \underbrace{\vec{\phi} \times \partial^\mu
  \vec{\phi}}_{{\vec V}^\mu_a} + \underbrace{\bar\Psi
\gamma^\mu \frac{\vec\tau}{2} \Psi}_{{\vec V}^\mu_b} 
+ \underbrace{\frac{g_A}{2f_\pi}\bar\Psi
\gamma^\mu \gamma_5 (\vec{\phi}\times\vec{\tau})\Psi}_{{\vec V}^\mu_c}
  \overbrace{-
\frac{1}{4f_\pi^2} \bar\Psi \gamma^\mu\left [
  \vec{\tau}\vec{\phi}^{\,2}-
  \vec{\phi}(\vec{\tau}\cdot\vec{\phi})\right]\Psi -
\frac{\vec{\phi}^{\,2}}{3f_\pi^2}(\vec{\phi}\times \partial^\mu
\vec{\phi})}^{{\vec V}^\mu_d} + {\cal O}(\frac{1}{f_\pi^3})
  \label{eq:vcurrent}
\end{equation}
\begin{equation}
{\vec A}^\mu = \underbrace{f_\pi \partial^\mu
  \vec{\phi}}_{{\vec A}^\mu_a} + \underbrace{g_A\bar\Psi
\gamma^\mu \gamma_5\frac{\vec\tau}{2} \Psi}_{{\vec A}^\mu_b} 
+ \underbrace{\frac{1}{2f_\pi}\bar\Psi
\gamma^\mu (\vec{\phi}\times\vec{\tau})\Psi}_{{\vec A}^\mu_c}
 + \overbrace{
\frac{2}{3f_\pi}\left[\vec{\phi} (\vec{\phi}\cdot\partial^\mu\vec{\phi})-
\vec{\phi}^{\,2}\partial^\mu\vec{\phi}\,\right]
-\frac{g_A}{4f_\pi^2} \bar\Psi \gamma^\mu\gamma_5\left [
  \vec{\tau}\vec{\phi}^{\,2}-
  \vec{\phi}(\vec{\tau}\cdot\vec{\phi})\right]\Psi }^{{\vec
  A}^\mu_d} + {\cal O}(\frac{1}{f_\pi^3}) \label{eq:acurrent}
\end{equation}
and determine the weak transition vertex where the $W-$boson is
absorbed.  These currents are, up to a factor, the hadronic
realization of the electroweak quark current $j^\mu_{cc+}$ for a system
of interacting pions and nucleons. Thus, ${\vec A}^\mu_a$ and ${\vec
V}^\mu_a$ account for the $W-$decay into one and two pions,
respectively, while ${\vec V}^\mu_b$ and ${\vec A}^\mu_b$ provide the
$WNN$ vector and axial vector couplings. Besides, ${\vec A}^\mu_c$ and
${\vec V}^\mu_c$ lead to contact $WNN\pi$ vertices and finally ${\vec
A}^\mu_d$ and ${\vec V}^\mu_d$ either contribute to processes with
more than one pion in the final state or provide loop corrections.

The overall normalization can be obtained, for instance, by relating
the currents of Eq.~(\ref{eq:vcurrent}) and Eq.~(\ref{eq:acurrent})
with the phenomenological vector and axial nucleon currents in the
$\langle N^\prime \pi | j^\mu_{\rm cc+}(0)| N \rangle$ matrix element
\begin{equation}
\langle p; \vec{p}^{~\prime}=\vec{p}+\vec{q}~ | j^\alpha_{cc+}(0) | n;
\vec{p}~\rangle =
\cos\theta_C\
\bar{u}(\vec{p}{~^\prime})(V^\alpha_N(q)-A^\alpha_N(q))u(\vec{p}\,) 
={\cal A}^\alpha
\label{eq:np}
\end{equation}
where the $u$'s are Dirac spinors for the neutron and proton,
normalized such that $\bar u u=1$, 
and vector and axial nucleon currents are given by
\begin{equation}
V_N^\alpha(q) = 2\times\left(F_1^V(q^2)\gamma^\alpha + {\rm
i}\mu_V \frac{F_2^V(q^2)}{2M}\sigma^{\alpha\nu}q_\nu\right), \qquad
A_N^\alpha (q)=  G_A(q^2) \times \left(\gamma^\alpha\gamma_5 + 
\frac{\slashchar{q}}{m_\pi^2-q^2}q^\alpha\gamma_5 \right) \, .
\label{eq:axial1} 
\end{equation}
We find\footnote{The $+1$ spherical component of a vector $\vec{A}$ is
defined as $A_{+1} = - \left(A_x + {\rm i}A_y\right)/\sqrt 2$.} that
$-\sqrt{2}\cos\theta_C\left ([V^\mu]_{+1} - [A^\mu]_{+1}\right)$
provides the $W^+$-absorption vertex, with the appropriate
normalization.

The magnetic part in Eq.~(\ref{eq:axial1}) is not provided
by the non-linear sigma model, which assumes structureless nucleons. 
We will improve on that by including the
$q^2$ dependence induced by the form factors in Eq.~(\ref{eq:axial1})
and adding the magnetic contribution, $F_2^V$ term, to the vector part
of the $W^+ N \to N$ amplitude.

For the nucleon vector form factors (FF) we use the parameterization of
Ref.~\cite{Galster:1971kv}
\begin{equation}
F_1^N = \frac{G_E^N+\kappa G_M^N}{1+\kappa}, \qquad \mu_N F_2^N =
\frac{G_M^N- G_E^N}{1+\kappa}, \quad G_E^p = \frac{G_M^p}{\mu_p}=
\frac{G_M^n}{\mu_n} = -(1+\lambda_n\kappa) \frac{G_E^n}{\mu_n \kappa} =
\left(\frac{1}{1-q^2/M^2_D}\right)^2 \label{eq:f1n}
\end{equation}
with $\kappa=-q^2/4M^2$, $M_D=0.843$ GeV, $\mu_p=2.792847$, $\mu_n=-1.913043$ and
$\lambda_n=5.6$.
\begin{equation}
 F_1^V(q^2) =  \frac12 \left (F_1^p(q^2)-F_1^n(q^2)\right),\qquad
 \mu_V F_2^V(q^2) = \frac12 \left ( \mu_p F_2^p(q^2) - \mu_n F_2^n(q^2)\right)\, . 
\end{equation}
The axial form factor is given by~\cite{Ericson:1988gk}
\begin{equation} 
G_A(q^2) = \frac{g_A}{(1-q^2/M_A^2)^2},\quad g_A=1.26, \quad M_A =
1.05~ {\rm GeV} \label{eq:axial}
\end{equation}

\begin{figure}\centering
  \includegraphics[width=0.9\textwidth]{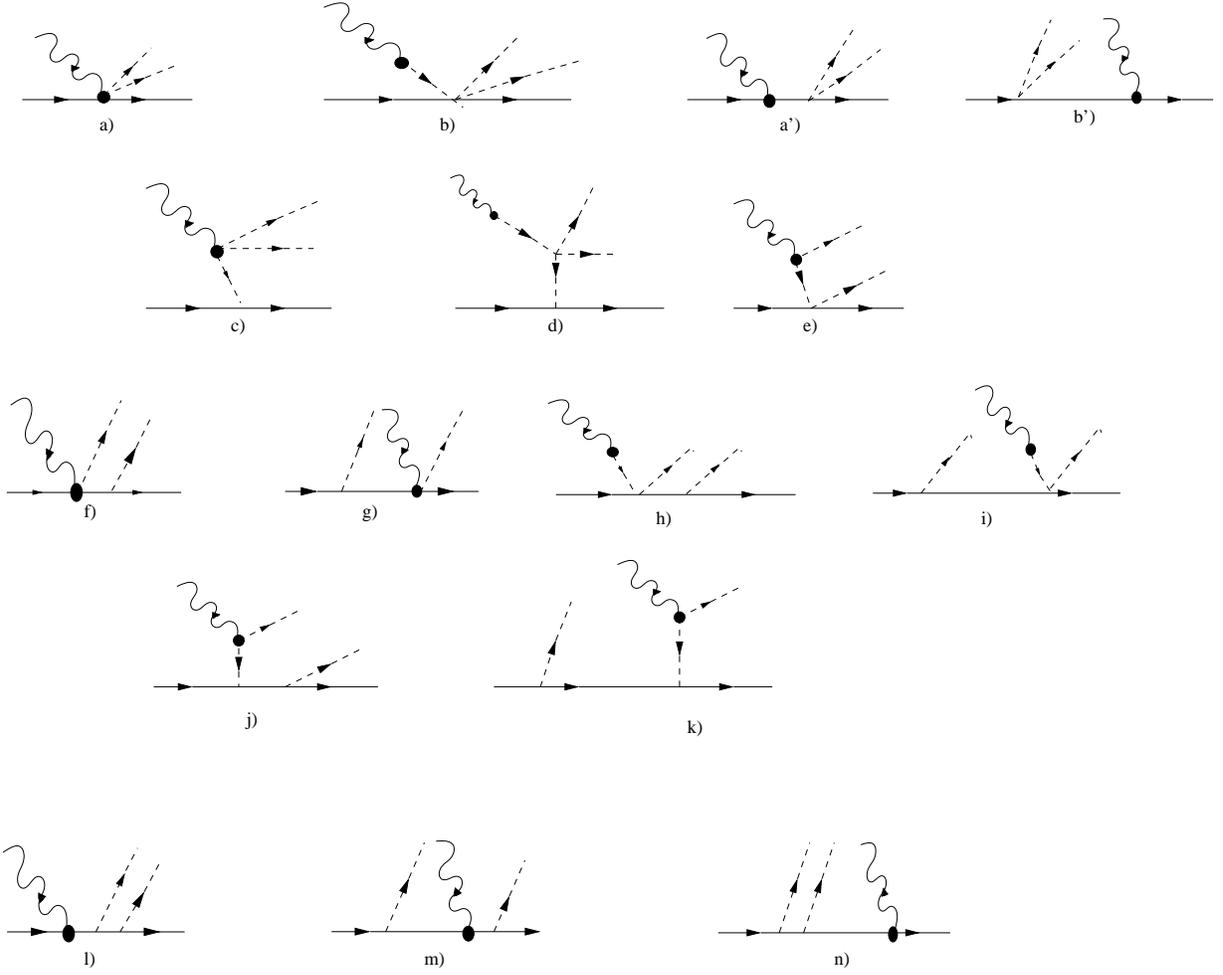}
  \caption{Nucleon pole, pion pole and contact terms contributing to
    $2\pi$ production.}
  \label{fig:fig1}
\end{figure}

Using this current we obtain the  sixteen
Feynman diagrams, depicted in Fig.~\ref{fig:fig1}, constructed
out of the $W^+N\to N$, $W^+N\to N\pi$, $W^+N\to N\pi\pi$, 
and the contact $W^+\pi\to \pi$ weak transition vertices
(Eqs.~(\ref{eq:vcurrent}--\ref{eq:acurrent})) and
the $\pi NN$, $\pi\pi NN$, $\pi\pi\pi NN$, $\pi\pi\pi\pi$ couplings
(Eqs.~(\ref{eq:lsigma}--\ref{eq:lint})).  Since we have
included a $q^2$ dependence ($F_1^V(q^2)$) on the Dirac part of the
vector $WNN$ vertex and to preserve vector current conservation, we
include the same form factor 
in the $V^\mu_{a,b,c,d}$ weak operators.

The expressions for the matrix elements of these terms, for all
possible channels, can be found in the appendix.

\subsection{Contribution of the $N^*(1440)$ resonance}
\label{sec:rop}

The Roper $N^*(1440)$ $P_{11}$ is the lowest lying baryon resonance
with an $s$-wave isoscalar two pion decay.  This suggests the possible
relevance of Roper excitation mechanisms in two pion production
processes close to threshold. Indeed, its importance has been clearly
established in the $\pi N\to\pi\pi N$ \cite{Oset:1985wt,Kamano:2004es}
and the $NN\to N\pi\pi$ \cite{Alvarez-Ruso:1997mx,Alvarez-Ruso:1998xg}
reactions where it plays a dominant role for certain isospin channels.
However, that is not the case for electromagnetically induced
reactions, due to the relatively weak coupling of the Roper to the
photons. See, for instance, Ref.~\cite{Gomez Tejedor:1993bq}.

\begin{figure}\centering
  \includegraphics[width=0.9\textwidth]{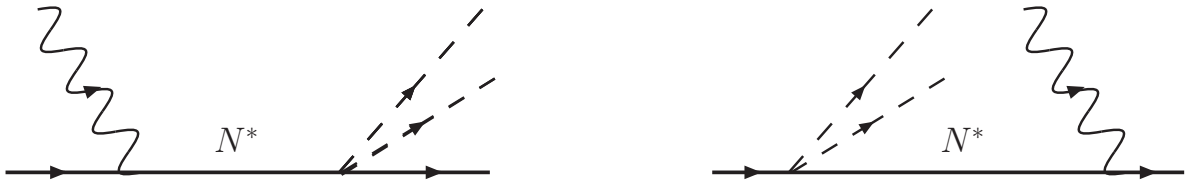}
  \caption{Direct  (left) and crossed  (right) Roper excitation  contributions to  $2\pi$ production.}
  \label{fig:rop}
\end{figure}

We include in our study the two mechanisms depicted in
Fig.\ref{fig:rop}, which account for the Roper production and its
decay into a nucleon and two pions in a $s$-wave isoscalar
state. However, we do not include the contribution of mechanisms in
which the Roper decays into $\pi\Delta (1232)$ states, because the two
produced pions are in $p$-wave and thus are not expected to be
relevant close to threshold.

\subsubsection{Roper interaction Lagrangians and currents}
The Lagrangian for the $s$-wave $N^*\to N\pi\pi$ decay can be written as
\begin{equation}
  {\cal L}_{N^*N\pi\pi} = 
  -c_1^*\frac{m_\pi^2}{f_\pi^2}\bar{\psi}_{N^*}\vec{\phi}^2\Psi +
   c_2^*\frac{1}{f_\pi^2}\bar{\psi}_{N^*}(\vec{\tau}\partial_0\vec{\phi})
   (\vec{\tau}\partial_0\vec{\phi})\Psi + h.c.\, ,
\label{eq:pipilag}
\end{equation}
neglecting terms of order ${\cal
O}(p^2/M^{*2})$~\cite{Bernard:1995gx}. Here, $\bar{\psi}_{N^*}$ is the
Roper field.  In Ref.~\cite{Alvarez-Ruso:1997mx} the best agreement
with $N N\to\pi\pi N N$ and the $\pi N\to\pi\pi N$ data was obtained
using $c_1^*=-7.27$ GeV$^{-1}$, $c_2^*=0$ GeV$^{-1}$.  This result was
obtained assuming a branching ratio of $7.5\%$ 
for the
$N(\pi\pi)^{I=0}_{J=0}$ decay mode and a total decay width of the
$N^*$, $\Gamma_{{\rm tot}} = 350$ MeV \cite{Eidelman:2004wy}.

The other required ingredient is the coupling of the Roper to the
charged weak current, that is the vertex $W^+n\to N^{*+}(1440)$. The
matrix elements can be written as
\begin{equation}
\langle N^{*+};\vec p_{*}=\vec p+\vec q\,|j^\alpha_{cc+}(0)|n;\vec p\,\rangle 
= \cos\theta_C\bar{u}_*(\vec p_{*}) J^\alpha_{cc*} u(\vec p\,) \, ,
\end{equation}
where $u_*$ is the Roper spinor and
\begin{equation}
J^\alpha_{cc*} = 
\frac{F_1^{V*}(q^2)}{\mu^2}(q^\alpha\slashchar{q}-q^2\gamma^\alpha)
+ i\frac{F_2^{V*}(q^2)}{\mu}\sigma^{\alpha\nu}q_\nu 
- G_A\gamma^\alpha\gamma_5 - \frac{G_P}{\mu}q^\alpha\slashchar q\gamma_5 
- \frac{G_T}{\mu} \sigma^{\alpha\nu}q_\nu\gamma_5
\end{equation}
is the most general form compatible with conservation of the vector
current and Dirac equation for the nucleon and Roper Dirac spinors.  A
factor $\mu = M + M_*$, with $M_*=1440$ MeV the mass of the Roper
resonance, is introduced in order to make the vector form factors
dimensionless.  The  $G_T$ term,
and unlike the elastic nucleon case
where it is zero due to G-parity invariance, 
does not need to vanish in the present case because the nucleon and the Roper
do not belong to the same isospin multiplet. Nevertheless, most analyses neglect its contribution
and we shall do so here.
% \cite{Meyer:2001js};

The $N^*N\pi$ coupling is described by the pseudovector Lagrangian
\begin{equation}
{\cal L}_{N^*N\pi} = \frac{\tilde{f}}{m_\pi}\bar{\Psi}_{N^*}\gamma^\mu\gamma_5
\vec{\tau}\cdot\partial_\mu\vec{\phi}\Psi\,+\, h.c.\, .
\label{eq:pseudovec}
\end{equation}
The decay width for this process is given by
\begin{equation}
\Gamma_{N^*\to\pi N}=\frac{3}{2\pi}\left(\frac{\tilde{f}}{m_\pi}\right)^2 
\frac{M}{W}|q_{\rm cm}|^3 \, ,
\label{eq:ropwidth}
\end{equation}
where $W$ is the $N^*$ invariant mass and $|q_{\rm cm}|$ is the
momentum of the outgoing pion in the outgoing $\pi N$ center of mass
frame.  Taking for this decay channel a branching ratio of $65\%$ and
the total width of the Roper $\Gamma = 350$
MeV~\cite{Eidelman:2004wy}, we get $\tilde{f}=0.48$.

\subsubsection{$N^*(1440)$ form factors}
Assuming  the pseudoscalar coupling
$G_P$ is dominated by the pion pole contribution, and imposing
partial conservation of the axial current (PCAC) hypothesis  we can relate
 $G_P$
with the axial coupling $G_A$ 
\begin{equation}
 G_P(q^2)  =\frac{\mu}{m_\pi^2-q^2}G_A(q^2) \, .
\label{eq:pcac}
\end{equation}
Furthermore  we can relate $G_A$ with the $N^*N\pi$ coupling
constant at $q^2=0$ using the  non-diagonal Goldberger--Treiman relation
\begin{equation}
G_A(0)= 2f_\pi \frac{\tilde{f}}{m_\pi}=0.63
\label{eq:gtnn*}
\end{equation}

The $q^2$ dependence of $G_A(q^2)$ is not constrained by theory so we
shall assume for it a dipole form of the type
\begin{equation}
  G_A(q^2) = \frac{G_A(0)}{(1-q^2/M_{A*}^2)^2} \, ,
\label{eq:axcoup}
\end{equation}
with an axial mass $M_{A*} = 1$ GeV.

The vector-isovector form factors $F_1^{V*}/\mu^2$ and $F_2^{V*}/\mu$ can be
related to the isovector part of the electromagnetic (EM) form
factors, $F_i^{V*}=F_i^{p*}-F_i^{n*}$, that can be determined from photo- and
electroproduction experiments.  The relevant experimental information
about these EM form factors is usually given in terms of helicity
amplitudes for the EM current $j_{{\rm e.m.}}^N$, defined\footnote{
Please note that the MAID group analysis \cite{Drechsel:2007if} define
$S_{1/2}^N$ with the opposite sign, which we take into account when
comparing to their data.} as~\cite{Warns:1989ie}
\begin{gather}
A_{1/2}^N = \sqrt{\frac{2\pi\alpha}{k_R}}\langle N^*\uparrow|
\sum_{{\rm pol}}
\epsilon\cdot 
j_{{\rm e.m.}}(0) |N\downarrow \rangle\,\xi \label{eq:heli0}\\
S_{1/2}^N = \sqrt{\frac{2\pi\alpha}{k_R}}\frac{|\vec{q}|}{\sqrt{-q^2}}
\langle N^*\uparrow|
\sum_{{\rm pol}}\epsilon\cdot j_{{\rm e.m.}}(0) |N\uparrow\rangle\,\xi \, ,
\label{eq:heli}
\end{gather}
where $N$ stands for proton or neutron, $\alpha=1/137$, $q$ is the
 momentum of the virtual photon, $k_R=(W^2-M^2)/2W$, with $W$ the
 energy of the Roper in its center of mass, and the polarization
 vectors are given by
\begin{equation}
\epsilon^{\pm}=\frac{1}{\sqrt{2}}(0,\mp 1,-i,0)\, ,
\label{eq:heli2}\end{equation}
and for a photon of momentum $q$ moving along the positive z-axis
\begin{equation}
\epsilon^{0}=\frac{1}{\sqrt{-q^2}}(|\vec{q}|,0,0,q^0)\, .
\label{eq:heli3}\end{equation}
The factor $\xi$ in Eqs. (\ref{eq:heli0}) and (\ref{eq:heli}) is given by the
relative sign between the $NN\pi$ and $N^*N\pi$ couplings~\cite{Warns:1989ie}
which we have taken to be positive (we will discuss this point later).

Finally, the EM  $\gamma N\to N^*$ current is written as
\begin{equation}
\langle N^*;\vec p_{*}=\vec p+\vec q\,|j_{{\rm e.m.}}^\alpha(0)|N;\vec p\,\rangle = \bar{u}_*(\vec p_{*})
\left[\frac{F_1^{N*}(q^2)}{\mu^2}(q^\alpha\slashchar{q}-q^2\gamma^\alpha)
+ i\frac{F_2^{N*}(q^2)}{\mu}\sigma^{\alpha\nu}q_\nu\right] u(\vec p\,)\, .
\label{eq:heli4}
\end{equation}
Using Eqs.~(\ref{eq:heli})-(\ref{eq:heli4}) we obtain the following relations:
\begin{gather}
A_{1/2}^N =|\vec{q}| g(q^2)\left[\frac{F_2^{N*}}{\mu}-\frac{q^2}{W+M} 
\frac{F_1^{N*}} {\mu^2}\right] \label{eq:heli5a}  \\
S_{1/2}^N = \frac{1}{\sqrt{2}}|\vec{q}|^2 g(q^2)
\left[\frac{F_1^{N*}}{\mu^2}-\frac{F_2^{N*}}{\mu} 
\frac{1}{W+M}  \right] \, ,
\label{eq:heli5b}
\end{gather}
with
\begin{equation}
  g(q^2)=\sqrt{\frac{8\pi\alpha(W+M)W^2}{M(W-M)((W+M)^2-q^2)}} \, .
\end{equation}
Inverting Eqs.~(\ref{eq:heli5a}) and (\ref{eq:heli5b}) we can obtain the
proton FF $F_i^{p*}$ as a function of the experimental helicities.
Unfortunately, there is no such information on the $\gamma n\to
N^{*0}$ transition so we need some theoretical assumptions to relate
the isovector helicity amplitudes with the EM ones. Following the
quark models predictions of \cite{Copley:1969qn} and
\cite{Close:1979bt} we shall assume $A^{n}_{1/2}=-2/3 A^p_{1/2}$ and
$S^{n}_{1/2}=0$. For a review of different models see
Ref.~\cite{AlvarezRuso:2003gj}.  Using these relations we can write
the neutron FF as a function of the proton ones and express the vector
FF in terms of only $F_1^{p*}$ and $F_2^{p*}$ as
\begin{gather}
F_1^{V*}=\frac{F_1^{p*}((M+W)^2-5 q^2/3)+2/3 F_2^{p*}(M+W) \mu}{(M+W)^2-q^2}\\
F_2^{V*}=\frac{F_2^{p*}(5(M+W)^2-3 q^2)\mu-2 F_1^{p*}q^2(M+W)}{3((M+W)^2-q^2)\mu}\, .
\end{gather}

We have fitted the proton-Roper EM transition form factors to the
experimental results for helicity amplitudes given in
\cite{Aznauryan:2004jd,Tiator:2003uu}
%and shown in fig.~\ref{fig:helix}. 
using a parameterization inspired by Lalakulich et al. \cite{Lalakulich:2006sw}
\begin{gather}
F_1^{p*}(q^2) = \frac{g_1^p/D_V}{1-q^2/X_1M_V^2} \, ,\\
F_2^{p*}(q^2) = \frac{g_2^p}{D_V}\left(1 - X_2\ln\left(1-\frac{q^2}{1\,{\rm GeV}^2}\right) \right)
\end{gather}
where $D_V= \left(1-q^2/M_V^2\right)^2$ with $M_V=0.84$ GeV.  
We have fitted the dimensionless parameters $g_1^p$,
$g_2^p$, $X_1$ and $X_2$ to the available experimental data, and found 
the following best fit (labeled as FF 1 in the results):
\begin{equation}
  g_1^p = -5.7 \pm 0.9  \, ,\quad g_2^p = -0.64 \pm 0.04 \, ,\quad 
  X_1 = 1.4 \pm 0.5 \, ,\quad X_2 = 2.47 \pm 0.12  \, ,
\label{eq:fit1}
\end{equation}
with a $\chi^2 / {\rm d.o.f} = 5.2$. Our best fit parameters $g_1^p$
and $X_1$ differ appreciably from those quoted in
Ref.~\cite{Lalakulich:2006sw}, in particular $g_1^p$ comes out with 
 opposite sign. This is because the authors of this
latter reference did not consider the existing extra minus sign among
their definition of $S_{1/2}^N$, which agrees with that of 
 Eq.~(\ref{eq:heli}) and used in this work, and the definition used in
 the MAID work. Indeed, if we do not consider this minus sign, we find
 values of the parameters in good agreement with those quoted in
 \cite{Lalakulich:2006sw}, with minor differences due to the use of
 different data sets. 

The Roper EM data have large error bars and it is possible to
accommodate quite different functional forms and values for these FF.
Thus, we shall compare other different models for the vector form
factors.  Firstly, we consider the constituent quark model with gluon,
pion and $\sigma$-meson exchange potentials as residual
interactions of Meyer et al.~\cite{Meyer:2001js} in set FF 2. In this
model the electromagnetic current included, in addition to the
one-body current, two-body exchange currents associated with the
quark-quark potentials. Here no assumption about the relation between
proton and nucleon form factor was assumed.  We will also consider
the recent parameterizations\footnote{Notice the opposite sign
convention for $F_V$ in \cite{Lalakulich:2006sw}} of Lalakulich et
al.~\cite{Lalakulich:2006sw} in the set labeled FF 3.  Note that this
work employed a pseudoscalar form for the $\pi NN$ coupling used in
deriving the Goldberger-Treiman relation, instead of the pseudovector,
Eq.~(\ref{eq:pseudovec}). 
We shall finally use the predictions of the
recent MAID analysis~\cite{Drechsel:2007if} in the set labeled FF 4.

From all this information we can now obtain definite expressions for the 
currents of the direct ($N^*d$) and crossed ($N^*c$) diagrams shown in Fig.~\ref{fig:rop}
\begin{gather}
{\cal A}_{N^*d}^\mu = 2g^*\bar{u}(\vec{p}\,')S_*(p+q)J^\mu_{cc*}(q)u(\vec{p})
\\
{\cal A}_{N^*c}^\mu = 2g^*\bar{u}(\vec{p}\,')\tilde{J}^\mu_{cc*}(q)S_*(p^\prime-q)
u(\vec{p}) \, , 
\end{gather}
with $g^* = c_1^*(m_\pi/f_\pi)^2$.
In the crossed diagram it appears $\tilde{J}^\mu_{cc*}$, that corresponds to
the crossed vertex $W^+N^*\to N$ and it is given by 
$\tilde{J}^\mu_{cc*} = \gamma^0(J^\mu_{cc*})^\dagger\gamma^0$.
We have here introduced the Roper propagator
\begin{equation}
S_*(p_*) = \frac{\slashchar{p}_*+M_*}{p_*^2-M_*^2 + 
i(M_*+W)\Gamma_{\rm tot}(W)/2} \, .
\label{eq:roprop}
\end{equation}
The total $W$-dependent decay width $\Gamma_{\rm tot}(W)$ includes all
the possible decay channels, namely $N\pi$, given by
Eq.~(\ref{eq:ropwidth}), $\pi\Delta$ and $N(\pi\pi)^{T=0}_{S=0}$.  For
these two cases we take
\begin{equation}
\Gamma_{\pi\Delta}(W) = 350\times 0.275 |p_\pi(W)|^3/|p_\pi(M_*)|^3\; {\rm MeV}
\end{equation}
and
\begin{equation}
\Gamma_{\pi\pi N} = 350\times 0.075 PhSp(W)/PhSp(M_*) \; {\rm MeV}\,,
\end{equation}
where $PhSp(W)$ is the phase space for the three body $\pi\pi N$ decay,
and $p_\pi$ is the pion momentum in the Roper rest frame.

The isospin structure of Eq.~(\ref{eq:pipilag}) determines that the
only allowed channels for the two pion s-wave decay of the Roper
resonance are $\pi^+\pi^-$ and $\pi^0\pi^0$. The above expressions are
given for the $\pi^+\pi^-$ channel, therefore we must include an
additional $1/2$ symmetry factor for the $\pi^0\pi^0$ channel.

\subsubsection{Relative sign 
between the Roper and non-resonant contributions}

Here we give some details on the used scheme to set up the 
relative signs between the different contributions:
\begin{itemize}
\item All relative phases in the non-resonant terms are completely
  fixed by the non-linear sigma model Lagrangian of
  Eqs.~(\ref{eq:lsigma}) and (\ref{eq:lint}), the currents deduced
  from it and shown in Eqs.~(\ref{eq:vcurrent}) and
  (\ref{eq:acurrent}), and the phenomenological $WNN$ vertex.

\item We have assumed the sign of the $N^*N\pi$ coupling to be the
  same as that of the $NN\pi$ coupling (positive). This choice fixes
  the global phases in Eqs. (\ref{eq:heli0}) and
  (\ref{eq:heli})~\cite{Warns:1989ie} and thus the phase of the vector
  part of the $WNN^*$ current. Besides the non-diagonal
  Goldberger-Treiman relation (Eq.(\ref{eq:gtnn*})) fixes the axial
  part of the $WNN^*$ current.

 \item Furthermore, once the relative sign of the $NN\pi$ and $N^*N\pi$ couplings has been set, 
 the study of the reactions $NN \to NN\pi\pi$  and $\pi N \to N \pi\pi$ in 
 Ref.~\cite{Alvarez-Ruso:1997mx} fixes the values for the $c_1^*$ 
 and $c_2^*$	 \ \ $N^*N\pi\pi$ couplings.

\end{itemize}
 
Taking the opposite sign for the $N^*N\pi$ coupling, does not affect
the results.  This is because the signs of both, the $WNN^*$ current
(see previous discussion), and the $c_1^*$ and $c_2^*$
couplings\footnote{These couplings would swap sign when the sign of
the $N^*N\pi$ coupling is changed, since they are fitted to the $NN
\to NN\pi\pi$ and $\pi N \to N \pi\pi$ data.} would change and
therefore the whole resonant contribution would not be affected at
all.

\section{Results}
\label{sec:res}

In this section, we present the results for all the two pion states
produced in the neutrino induced reactions on nucleon targets,
i.e. $\nu p\to \mu^- p \pi^+\pi^0$, $\nu p\to \mu^- n \pi^+\pi^+$,
$\nu n\to \mu^- p \pi^+\pi^-$, $\nu n\to \mu^- n \pi^+\pi^0$ and $\nu
n\to \mu^- p \pi^0\pi^0$.

Our model includes all relevant terms for neutrino induced two pion
production close to threshold.  In addition to the contribution of
nucleon and pion pole and contact terms described by the chiral
Lagrangian the contribution of the $N^*(1440)$ resonance coupling to
two s-wave pions is included. The $\Delta(1232)$ and other higher
resonances are not considered as their contributions would vanish at
threshold.

\begin{figure}
 \includegraphics[width=0.8\textwidth]{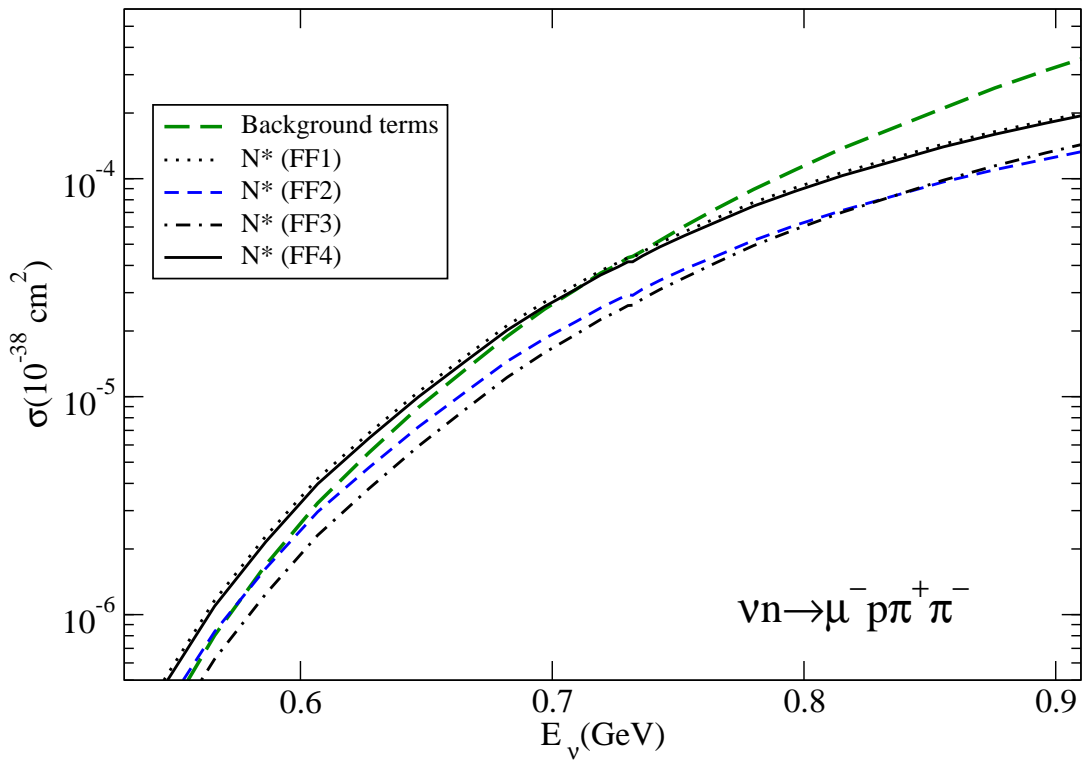}
\caption{Cross section for the $\nu n\to \mu^- p \pi^+\pi^-$ reaction as a function of the neutrino energy. The 
interference between background and the $N^*$ contribution is not shown. See text for details.}
  \label{fig:3}     
\end{figure} 
In Fig.~\ref{fig:3}, we present the results for the cross section for
the process $\nu n\to \mu^- p \pi^+\pi^-$.  We show separately the
contribution of the background terms coming from the nucleon pole,
pion pole and contact terms as well as the contribution of the Roper
resonance as calculated by using the various form factors described in
section~\ref{sec:CC}. The interference between background and the
Roper contribution is not shown.  We see that the background terms
dominate the cross section for neutrino energies $E_\nu>0.7$ GeV. At
lower energies the contribution from the Roper could be larger or
smaller than the background depending upon the vector form factors
used for the $W^+NN^*$ transition. The parameterization determined by
the recent MAID analysis~\cite{Drechsel:2007if} gives the largest
Roper contribution to the cross section.  The differences in the
predictions for the cross sections using the various parameterizations
could reach a factor two.  The Roper contribution is specially
sensitive to $F_2^{V*}(q^2)$ which is negative in contrast to the
positive value which one gets in the case of the nucleon. This has
interesting implications for the vector--axial vector interference
contribution for the cases of $\nu$ and $\bar{\nu}$ excitation of the
$N^*(1440)$. The comparison of data on $\nu$ and $\bar{\nu}$ induced
two pion production in this channel, if available in the future, could
be used to study the $W N N^*$ transition and better constrain the
corresponding form factors.
\begin{figure}
 \includegraphics[width=0.8\textwidth]{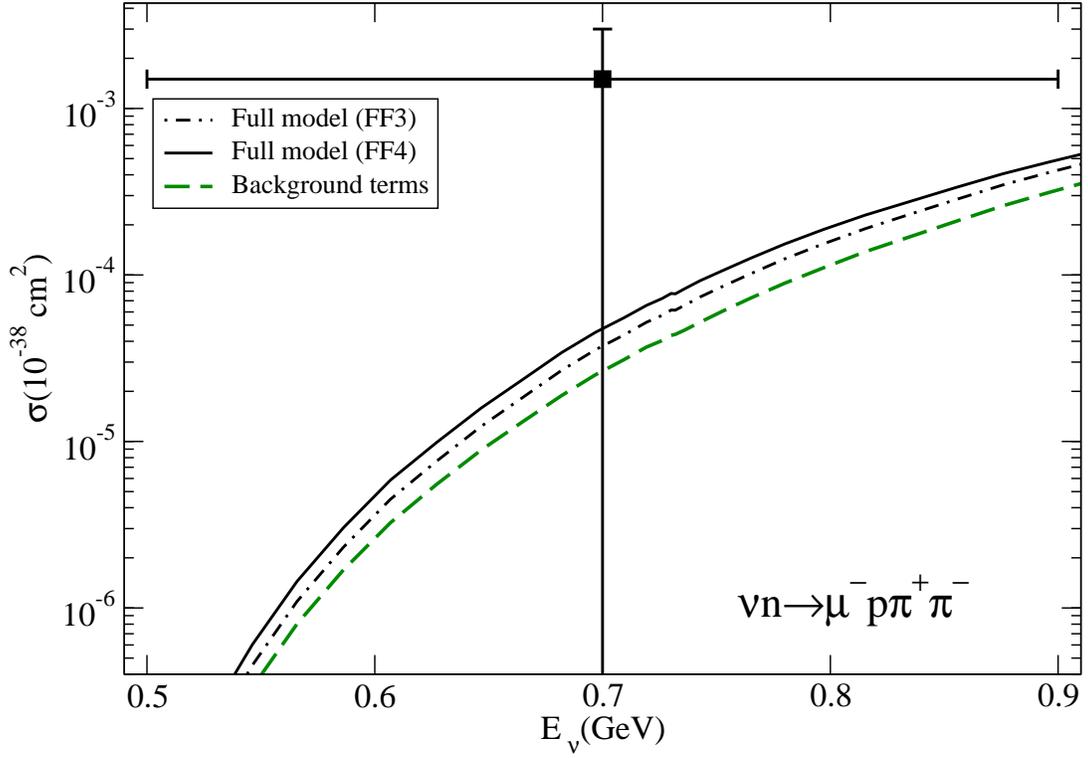}
\caption{Cross section for the $\nu n\to \mu^- p \pi^+\pi^-$ reaction
as a function of the neutrino energy.   Data from
Ref.~\cite{Kitagaki:1986ct}. }
  \label{fig:4}     
\end{figure}

In Fig.~\ref{fig:4}, we compare our results for the same process with
experiment.  The largest results are obtained with the set 4, and the
smallest from the set 3 of nucleon-Roper transition form factors.  The other two sets give cross sections, not
shown in the figure, lying between these two limits. Our results with
the full model are larger than those obtained using only the
background terms as done by Adjei et al. \cite{Adjei:1980nj}, however
the results are still lower than the central value of the experiment.
Experimental data with better statistics is highly desirable in this
channel to make a decisive comparision with the predictions of our
model.

We expect our model to have a limited region of applicability from
threshold to an invariant mass of the $\pi\pi N$ system below $1.4$GeV. 
For muon neutrinos, that implies a LAB energy under 750 MeV,
beyond which some additional contributions from the $\Delta(1232)$ and
other higher resonances will become relevant.  In order to make a
meaningful comparison with the theoretical calculations for threshold
two pion production, Adjei et al. \cite{Adjei:1980nj} suggested a
kinematical cut on phase space which was implemented in the
experimental analysis made by Day et al. \cite{Day:1984nf} and
Kitagaki et al. \cite{Kitagaki:1986ct}.
%In Ref.~\cite{Adjei:1980nj} some kinematics cuts in the outgoing
%hadrons phase space were introduced in order to extend the range of
%validity of the chiral model to high incident neutrino energies.
These cuts were defined as
\begin{gather}
q_\pi^2 \leq \left((1 + \eta/2)m_\pi\right)^2 \, ,\\
p\cdot q_\pi \leq (M + (1+\eta)m_\pi)^2 - M^2 -m_\pi^2 \,  \\
p^\prime\cdot q_\pi \leq (M + (1+\eta)m_\pi)^2 - M^2 -m_\pi^2 \, ,
\label{eq:cuts}
\end{gather}
with $q_\pi = (k_{\pi_1} + k_{\pi_2})/2$.  Different choices for
$\eta$, specifically $\eta = 1/4$, $2/4$, $3/4$, were proposed.  As
explained in Ref.~\cite{Adjei:1980nj}, the first inequality keeps the
individual pion momenta close to the average pion momentum, while the
last two restrict the phase space to regions of small invariant mass
of the three hadrons.

\begin{figure}
 \includegraphics[width=0.8\textwidth]{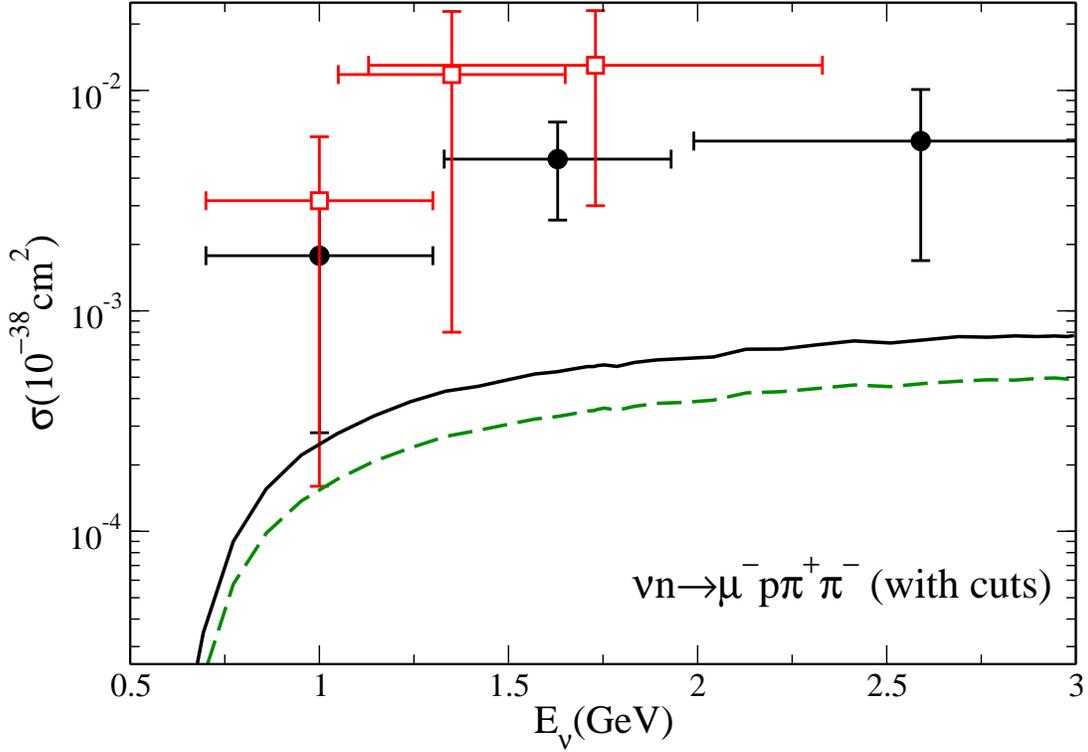}
\caption{Cross section for the $\nu n\to \mu^- p \pi^+\pi^-$ with cuts
as explained in the text.  Dashed line: Background terms. Solid line:
Full model with set 1 of nucleon-Roper transition FF.  Data from
Ref.~\cite{Kitagaki:1986ct} (solid circles) and Ref.~\cite{Day:1984nf}
(open squares).}
\label{fig:5}     
\end{figure}
Using these kinematic restrictions on the phase space with $\eta=3/4$,
as used by Refs.~\cite{Day:1984nf,Kitagaki:1986ct}, we present the
results for the cross section for the $\nu n\to \mu^- p \pi^+\pi^-$
channel in Fig.~\ref{fig:5} and compare with their data. We show our
results with only background terms and with the full model evaluated
using the set 1 of nucleon-Roper transition form factors. Other sets
give a similar result in this case.  We find that, even in this
kinematic region, the theoretical results including the resonance
contribution are lower than the experimental.

\begin{figure}
  \includegraphics[width=0.8\textwidth]{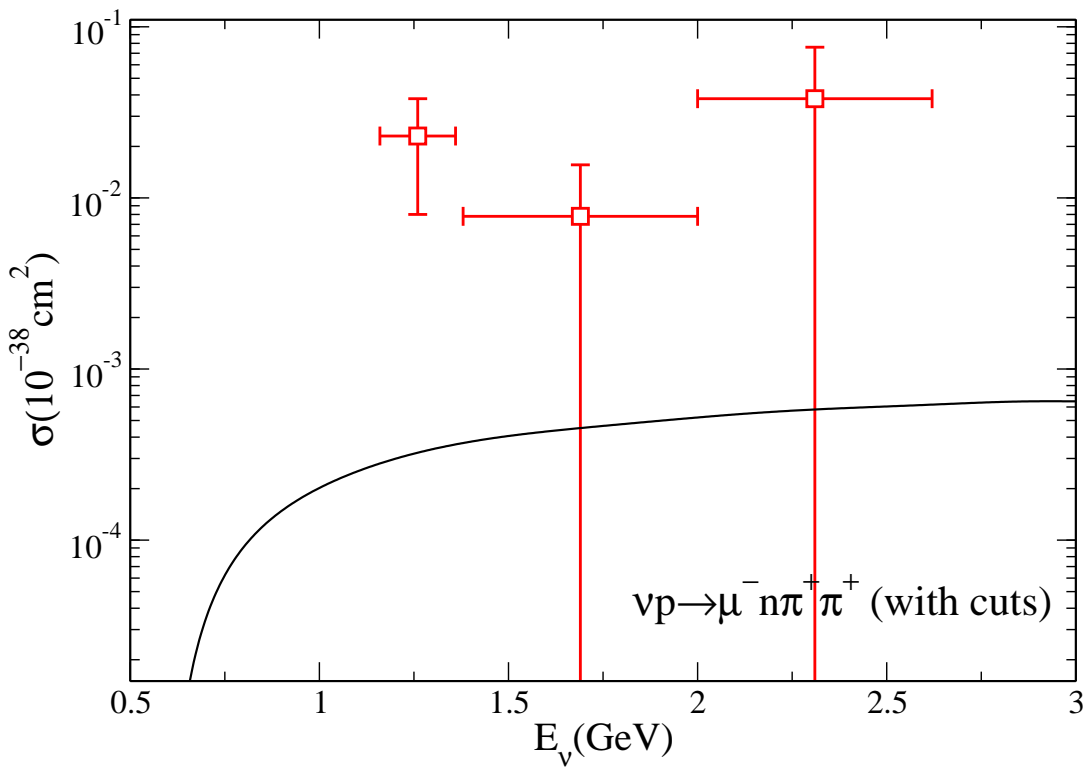}
  \caption{Cross section for the $\nu p\to \mu^- n \pi^+\pi^+$ with cuts as 
explained in the text. Note that there are no contributions from 
the $N^*(1440)$ resonance to this channel. Data from Ref.~\cite{Day:1984nf}.}
\label{fig:6}     
\end{figure}

In Fig.~\ref{fig:6}, we present the results, with the same cuts, for
the total cross section for the channel $\nu p\to \mu^- n \pi^+\pi^+$
and compare with the data of Ref.~\cite{Day:1984nf}.  For this channel
there is no contribution from the $N^*(1440)$ resonance. Our results
are in agreement with those of Adjei et al. \cite{Adjei:1980nj} and
are consistent, within errors, with the experiment in the higher
energy region.  However, we underestimate the first point
($E_\nu=1.25$ GeV) by more than one order of magnitude.  This
disagreement may be ascribed to the low statistics of the experiment
after the kinematical cuts have been implemented~\cite{Day:1984nf}, to
inadvertant inclusion of some P wave pions while implementing the
kinematical cuts in the experimental analysis but also to possible
additional reaction mechanisms not included in our model.  Note that
the kinematics of the plot explore the region just up to $1.4$ GeV of
final hadron state invariant mass. At this energy we do not expect our
threshold production model to account for all the mechanisms of pion
production, many of which would be in $p$-wave and therefore beyond
the scope of our work. This mechanisms would include $\Delta$
resonance terms similar to those studied in \cite{Gomez
Tejedor:1993bq} in the context of photo-induced reactions. In any
case, the disagreement at $E=1.25$ GeV underscores the need for better
experimental data in this channel around neutrino energy $E=1$ GeV.

\begin{figure}
\includegraphics[width=0.8\textwidth]{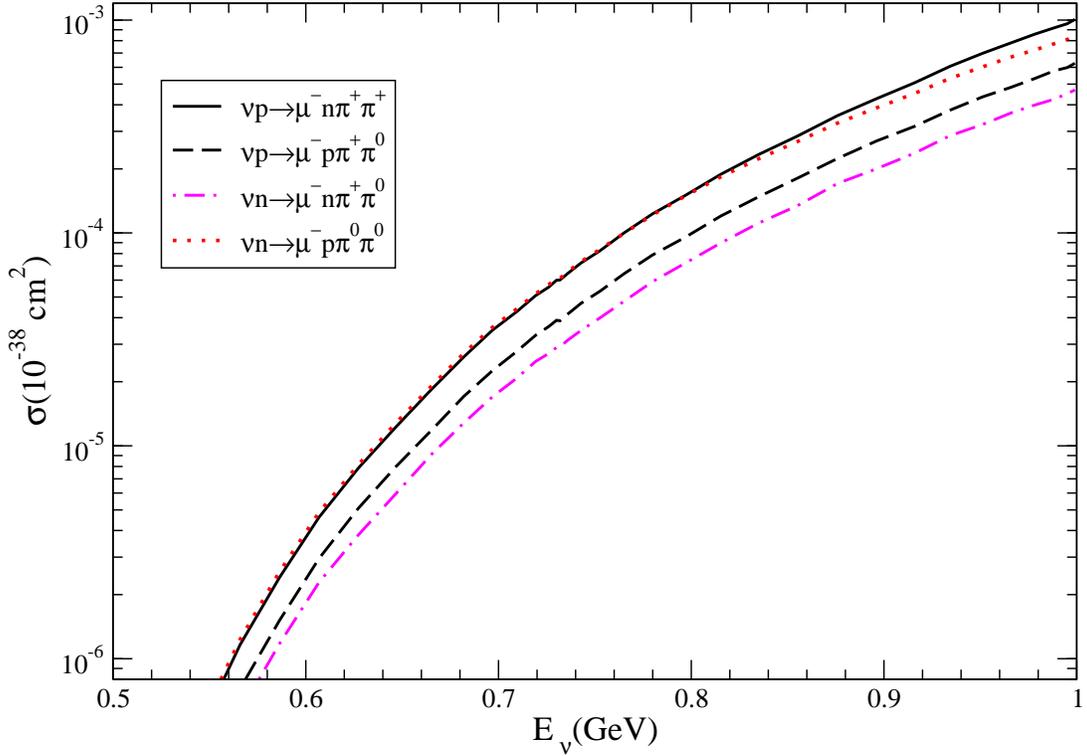}
\caption{Cross sections as a function of the neutrino energy.  All
calculations correspond to the full model with the FF1 set of nucleon-Roper
transition form factors.}
  \label{fig:7}     
\end{figure}

Finally, in Fig.~\ref{fig:7}, we show our results for the threshold
production of two pions in the other channels. Again we only show our
results with only background terms and with the full model evaluated
using the set 1 of nucleon-Roper transition form factors. The use of
different sets gives  similar results for the full model
calculation. There is no data in the low energy region; the limited
data available in these channels at higher energies have not been
analyzed in the kinematical region of the threshold production of two
pions and thus a comparison with our present results is not possible.

To summarize, we have studied the threshold production of two pions
induced by neutrinos on nucleon targets using a chiral Lagrangian to
calculate the nucleon pole, pion pole and contact terms. The
contribution from the excitation of the Roper resonance, $N^*(1440)$
has been included. The vector transition form factors are determined
from the available data on the helicity amplitudes of its
electromagnetic excitation. The axial form factors are obtained using
PCAC and the experimental data on $N^*N\pi$ decay. The $q^2$
dependence of the axial form factor has been assumed to be of the
dipole form. It is found that the Roper resonance contributes strongly
to the $\nu n\to \mu^- p \pi^+\pi^-$ and $\nu n\to \mu^- p \pi^0\pi^0$
channels. Its contribution is as important as the contribution of the
nucleon, pion pole and contact terms, up to energies of $E_\nu\sim
700$ MeV. The resonance contribution is sensitive to the form factors
used for the $WNN^*$ transition, and the experimental data, if
available with better statistics for the $\nu n\to \mu^- p \pi^+\pi^-$
and $\nu n\to \mu^- p \pi^0\pi^0$ channels, could be used for the
determination of these form factors. For other channels, as well as
for these, the theoretical results underestimate the experimental
results.  Furthermore, physical channels involving $\Delta$ production
could become relevant at beam energies higher than $750$ MeV, where
the threshold approximation we have assumed in our model would no
longer be valid.  Nevertheless, the low statistical significance of
data does not provide a conclusive test of our model.  Availability of
improved data, specially from future experiments on neutrino induced
two pion production in the region of energies $E_\nu < 1$ GeV will be
very useful in the study of the electroweak properties of the
$N^*(1440)$ resonance.

\begin{acknowledgments}
We acknowledge discussions with L. Alvarez Ruso and
L. Tiator. M.V. acknowledges the hospitality of the Department of
F\'\i sica Te\'orica and IFIC where part of this work was done.  This
work was partially supported by MEC contracts FPA2004-05616,
FIS2005-00810, FIS2006-03438 and FPA2007-65748, and by the Spanish Consolider-Ingenio 2010 Programme CPAN (CSD2007-00042), by the Generalitat
Valenciana contract ACOMP07/302, by Junta de Andaluc\'\i a and Junta
de Castilla y Le\'on under contracts FQM0225 and SA016A07 and by the
EU Integrated Infrastructure Initiative Hadron Physics Project
contract RII3-CT-2004-506078.
\end{acknowledgments}

%%%%%%%%%%%%%%%%%%%%%%%%%%%%%%%%%%%%%%%%%%%%%%%%%%%%%%%%%%%%%%%%%%%%%

\appendix

\section{Hadronic current for the different charge channels}
\label{sec:curr}

The contributions are labelled following  Fig. \ref{fig:fig1}.
We give explicit expressions for all channels except for $W^+n\to p\pi^0\pi^0$
which can be obtained via the following isospin relation
\begin{equation}
{\cal A}(W^+n\to p\pi^0\pi^0)={\cal A}(W^+n\to p\pi^+\pi^-)
-\frac{1}{\sqrt2}\left({\cal A}(W^+p\to p\pi^+\pi^0)-
{\cal A}(W^+n\to n\pi^+\pi^0)
\right).
\end{equation}

In all these expresions we have used for the nucleon propagator
\begin{equation}
S(p) = \frac{\slashchar{p}+M}{p^2-M^2+i\epsilon} \, 
\end{equation}
and $V^\mu - A^\mu$ are defined as in Eq.~(\ref{eq:axial1}).

\subsection{ $W^+_qp_{p}\to p_{p'}\pi^+_{k_1}\pi^0_{k_2}$}

\begin{equation}
{\cal A}^\mu_a=\frac{1}{2\sqrt2f_\pi^2}\cos\theta_c\ \overline{u}(\vec{p}\,')
\left(2F_1^V\gamma^\mu-g_A\,\gamma^\mu\gamma_5
\right)
u(\vec{p})
\end{equation}
\begin{equation}
{\cal A}^\mu_b=\frac{g_A}{6\sqrt2f_\pi^2}\cos\theta_c\ \frac{q^\mu}{q^2-m_\pi^2}\
\overline{u}(\vec{p}\,')
\left(3\slashchar{k}_2+2M\,
\right)\gamma_5\,
u(\vec{p})
\end{equation}

\begin{equation}
{\cal A}^\mu_c=\frac{g_A\sqrt2}{3f_\pi^2}\cos\theta_c\ \frac{q^\mu-3k_1^\mu}{(q-k_1-k_2)^2-m_\pi^2}\
2M\,\overline{u}(\vec{p}\,')
\gamma_5\,
u(\vec{p})
\end{equation}
\begin{equation}
{\cal A}^\mu_d=\frac{g_A}{3\sqrt2f_\pi^2}\cos\theta_c\ \frac{q^\mu}{q^2-m_\pi^2}\
\frac{4qk_1-2qk_2+2k_1k_2-q^2}{(q-k_1-k_2)^2-m_\pi^2}\
2M\,\overline{u}(\vec{p}\,')
\gamma_5\,
u(\vec{p})
\end{equation}

\begin{equation}
{\cal A}^\mu_e=\frac{1}{\sqrt2f_\pi^2}\cos\theta_c\ \frac{2k_2^\mu-q^\mu}{(q-k_2)^2-m_\pi^2}\
2F_1^V\,\overline{u}(\vec{p}\,')
\slashchar{k}_1\
\,u(\vec{p})
\end{equation}
\begin{equation}
{\cal A}^\mu_{b'} = -\frac{\sqrt2}{4f_\pi^2}\cos\theta_c\ 
\overline{u}(\vec{p}\,')\left(V^\mu - A^\mu\right) S(p'-q)\left(
\slashchar{k}_1-\slashchar{k}_2\right)
u(\vec{p})
\end{equation}
\begin{equation}
{\cal A}^\mu_f=-\frac{g_A}{2\sqrt2 f_\pi^2}\cos\theta_c\ 
\overline{u}(\vec{p}\,')\left(
\slashchar{k}_2\gamma_5S(p'+k_2)
\,2F_1^Vg_A\gamma^\mu\gamma_5-\slashchar{k}_2\
\gamma_5S(p'+k_2)\gamma^\mu\,\right)u(\vec{p})
\end{equation}
\begin{eqnarray}
{\cal A}^\mu_g=&&\frac{g_A}{2\sqrt2 f_\pi^2}\cos\theta_c\left\{
\overline{u}(\vec{p}\,')2F_1^Vg_A\gamma^\mu\gamma_5
\left(-S(p-k_2)\slashchar{k}_2\gamma_5+
2S(p-k_1)\slashchar{k}_1\gamma_5
\right)
u(\vec{p})\right.\nonumber\\
&&\left.-\overline{u}(\vec{p}\,')\gamma^\mu
\left(-\,S(p-k_2)\slashchar{k}_2\gamma_5+
2\,S(p-k_1)\slashchar{k}_1\gamma_5
\right)
u(\vec{p})\right\}
\end{eqnarray}
\begin{equation}
{\cal A}^\mu_h=-\frac{g_A}{4\sqrt2f_\pi^2}\cos\theta_c\ \frac{q^\mu}{q^2-m_\pi^2}\
\overline{u}(\vec{p}\,')\,\slashchar{k}_2\gamma_5 S(p'+k_2)
(\slashchar{q}+\slashchar{k}_1)
u(\vec{p})
\end{equation}
\begin{equation}
{\cal A}^\mu_i=-\frac{g_A}{4\sqrt2f_\pi^2}\cos\theta_c\ \frac{q^\mu}{q^2-m_\pi^2}\
\overline{u}(\vec{p}\,')\ \left(
\left(\slashchar{q}+\slashchar{k}_1\right)S(p-k_2)
\slashchar{k}_2\gamma_5-2
(\slashchar{q}+\slashchar{k}_2)S(p-k_1)
\slashchar{k}_1\gamma_5\right)\
u(\vec{p})
\end{equation}
\begin{equation}
{\cal A}^\mu_j=\frac{g_A^2}{2\sqrt2f_\pi^2}\cos\theta_c\ 2F_1^V\frac{q^\mu-2k_1^\mu}{(q-k_1)^2-m_\pi^2}\
\overline{u}(\vec{p}\,')\slashchar{k}_2\gamma_5
S(p'+k_2)(\slashchar{q}-\slashchar{k}_1)\gamma_5\,
u(\vec{p})
\end{equation}
\begin{eqnarray}
{\cal A}^\mu_k=\frac{g_A^2}{2\sqrt2f_\pi^2}\cos\theta_c\ 
2F_1^V\bigg(&&
\frac{q^\mu-2k_1^\mu}{(q-k_1)^2-m_\pi^2}\
\overline{u}(\vec{p}\,')(\slashchar{q}-\slashchar{k}_1)\gamma_5
S(p-k_2)\slashchar{k}_2\gamma_5\,
u(\vec{p})\nonumber\\
&&-2\frac{q^\mu-2k_2^\mu}{(q-k_2)^2-m_\pi^2}\
\overline{u}(\vec{p}\,')(\slashchar{q}-\slashchar{k}_2)\gamma_5
S(p-k_1)\slashchar{k}_1\gamma_5\,
u(\vec{p})
\bigg)
\end{eqnarray}
\begin{equation}
{\cal A}^\mu_m = -\frac{g_A^2}{2\sqrt2f_\pi^2}\cos\theta_c\ 
\overline{u}(\vec{p}\,')\slashchar{k}_2\gamma_5S(p'+k_2)
\left( V^\mu-A^\mu \right)
S(p-k_1) \slashchar{k}_1\gamma_5\,
u(\vec{p})
\end{equation}
\begin{equation}
{\cal A}^\mu_n=\frac{g_A^2}{2\sqrt2f_\pi^2}\cos\theta_c\ 
\overline{u}(\vec{p}\,') \left(V^\mu - A^\mu\right)
S(p-k_1-k_2)\left(\slashchar{k}_2\gamma_5S(p-k_1)\slashchar{k}_1\gamma_5
-\slashchar{k}_1\gamma_5S(p-k_2)\slashchar{k}_2\gamma_5\right)u(\vec{p})
\end{equation}

\subsection{$W^+_qp_p\to n_{p'}\pi^+_{k_1}\pi^+_{k_2}$} 
\begin{equation}
{\cal A}^\mu_a=\frac{1}{f_\pi^2}\cos\theta_c\ \overline{u}(\vec{p}\,')
\left(2F_1^V\,\gamma^\mu-g_A\,\gamma^\mu\gamma_5
\right)
u(\vec{p})
\end{equation}
\begin{equation}
{\cal A}^\mu_b=\frac{g_A}{6f_\pi^2}\cos\theta_c\ \frac{q^\mu}{q^2-m_\pi^2}\
\overline{u}(\vec{p}\,')
\left( 3 \slashchar{k}_1+3 \slashchar{k}_2
+ 4M\right)\gamma_5\,
u(\vec{p})
\end{equation}
\begin{equation}
{\cal A}^\mu_c=\frac{2g_A}{3f_\pi^2}\cos\theta_c\ \frac{(q^\mu-3k_1^\mu)+(q^\mu-3k_2^\mu)}
{(q-k_1-k_2)^2-m_\pi^2}\
2M\,\overline{u}(\vec{p}\,')
\gamma_5\,
u(\vec{p})
\end{equation}
\begin{equation}
{\cal A}^\mu_d=\frac{g_A}{3f_\pi^2}\cos\theta_c\ \frac{q^\mu}{q^2-m_\pi^2}\ 
\left\{
\frac{(4qk_1-2qk_2+2k_1k_2-q^2)}{(q-k_1-k_2)^2-m_\pi^2}
+ \frac{(4qk_2-2qk_1+2k_1k_2-q^2)}{(q-k_1-k_2)^2-m_\pi^2}\right\}\
2M\,\overline{u}(\vec{p}\,')
\gamma_5\,
u(\vec{p})
\end{equation}
\begin{equation}
{\cal A}^\mu_e=2F_1^V\, \frac{1}{f_\pi^2}\cos\theta_c\ 
\left\{\frac{2k_1^\mu-q^\mu}{(q-k_1)^2-m_\pi^2}
\overline{u}(\vec{p}\,')\slashchar{k}_2 u(\vec{p}) +
\frac{2k_2^\mu-q^\mu}{(q-k_2)^2-m_\pi^2}
\overline{u}(\vec{p}\,')\slashchar{k}_1u(\vec{p})
\right\}
\end{equation}
\begin{eqnarray}
{\cal A}^\mu_f=&&-\frac{g_A}{2 f_\pi^2}\cos\theta_c\left\{\ \
\overline{u}(\vec{p}\,')
\left(\slashchar{k}_1\gamma_5S(p'+k_1)
+\slashchar{k}_2\
\gamma_5S(p'+k_2)\right)2F_1^Vg_A\gamma^\mu\gamma_5\,u(\vec{p})\right.\nonumber\\
&&\left.-\overline{u}(\vec{p}\,')\left(\slashchar{k}_1\gamma_5S(p'+k_1)
+\slashchar{k}_2\
\gamma_5S(p'+k_2)\right)\gamma^\mu u(\vec{p})\right\}
\end{eqnarray}
\begin{eqnarray}
{\cal A}^\mu_g=&&\frac{g_A}{2 f_\pi^2}\cos\theta_c\left\{\ \
\overline{u}(\vec{p}\,')2F_1^Vg_A\gamma^\mu\gamma_5
\left(S(p-k_1)\slashchar{k}_1\gamma_5+
S(p-k_2)\slashchar{k}_2\gamma_5
\right)
u(\vec{p})\right.\nonumber\\
&&\left.-\overline{u}(\vec{p}\,')\gamma^\mu
\left(S(p-k_1)\slashchar{k}_1\gamma_5+
\,S(p-k_2)\slashchar{k}_2\gamma_5
\right)
u(\vec{p})\right\}
\end{eqnarray}
\begin{equation}
{\cal A}^\mu_h=-\frac{g_A}{4f_\pi^2}\cos\theta_c\ \frac{q^\mu}{q^2-m_\pi^2}\
\overline{u}(\vec{p}\,')\left(\slashchar{k}_1\gamma_5 S(p'+k_1)
\left(\slashchar{q}+\slashchar{k}_2\right)
+\slashchar{k}_2\gamma_5 S\left(p'+k_2\right)
\left(\slashchar{q}+\slashchar{k}_1\right)\right)u(\vec{p})
\end{equation}
\begin{equation}
{\cal A}^\mu_i=\frac{g_A}{4f_\pi^2}\cos\theta_c\ \frac{q^\mu}{q^2-m_\pi^2}\
\overline{u}(\vec{p}\,')\left(
\left(\slashchar{q}+\slashchar{k}_1\right)S(p-k_2)
\slashchar{k}_2\gamma_5
+(\slashchar{q}+\slashchar{k}_2)S(p-k_1)
\slashchar{k}_1\gamma_5\right)
u(\vec{p})
\end{equation}
\begin{multline}
{\cal A}^\mu_j=\frac{g_A^2}{2f_\pi^2}\cos\theta_c2 
F_1^V\left(\frac{q^\mu-2k_1^\mu}{(q-k_1)^2-m_\pi^2}\
\overline{u}(\vec{p}\,')\slashchar{k}_2\gamma_5
S(p'+k_2)(\slashchar{q}-\slashchar{k}_1)\gamma_5
u(\vec{p}) \right.\\\left.
+\frac{q^\mu-2k_2^\mu}{(q-k_2)^2-m_\pi^2}\
\overline{u}(\vec{p}\,')\slashchar{k}_1\gamma_5
S(p'+k_1)(\slashchar{q}-\slashchar{k}_2)\gamma_5%\right)
u(\vec{p}) \right)
\end{multline}
\begin{multline}
{\cal A}^\mu_k=-\frac{g_A^2}{2f_\pi^2}\cos\theta_c 2F_1^V 
\left(\frac{q^\mu-2k_1^\mu}{(q-k_1)^2-m_\pi^2}\
\overline{u}(\vec{p}\,')(\slashchar{q}-\slashchar{k}_1)\gamma_5
S(p-k_2)\slashchar{k}_2\gamma_5\,u(\vec{p})
\right.\\\left.
+\frac{q^\mu-2k_2^\mu}{(q-k_2)^2-m_\pi^2}\
\overline{u}(\vec{p}\,')(\slashchar{q}-\slashchar{k}_2)\gamma_5
S(p-k_1)\slashchar{k}_1\gamma_5\,
u(\vec{p})\right)
\end{multline}
\begin{equation}
{\cal A}^\mu_m=-\frac{g_A^2}{2f_\pi^2}\cos\theta_c\ 
\overline{u}(\vec{p}\,')\left(\slashchar{k}_1\gamma_5S(p'+k_1)
\left( V^\mu - A^\mu \right)
S(p-k_2)\slashchar{k}_2\gamma_5%\right.\\\left.
+\slashchar{k}_2\gamma_5S(p'+k_2)
\left( V^\mu - A^\mu \right)
S(p-k_1)\slashchar{k}_1\gamma_5\right)u(\vec{p})
\end{equation}
\subsection{$W^+_qn_p\to p_{p'}\pi^+_{k_1}\pi^-_{k_2}$}
\begin{equation}
{\cal A}^\mu_a=-\frac{1}{2f_\pi^2}\cos\theta_c\ \overline{u}(\vec{p}\,')
\left(2F_1^V\,\gamma^\mu-g_A\,\gamma^\mu\gamma_5
\right)
u(\vec{p})
\end{equation}
\begin{equation}
{\cal A}^\mu_b=-\frac{g_A}{6f_\pi^2}\cos\theta_c\ \frac{q^\mu}{q^2-m_\pi^2}\
\overline{u}(\vec{p}\,')
\left(3\slashchar{k}_1 +2M\,\right)\gamma_5\,
u(\vec{p})
\end{equation}
\begin{equation}
{\cal A}^\mu_c=-\frac{2g_A}{3f_\pi^2}\cos\theta_c\ \frac{q^\mu-3k_2^\mu}
{(q-k_1-k_2)^2-m_\pi^2}\
2M\,\overline{u}(\vec{p}\,')
\gamma_5\,
u(\vec{p})
\end{equation}
\begin{equation}
{\cal A}^\mu_d=-\frac{g_A}{3f_\pi^2}\cos\theta_c\ \frac{q^\mu}{q^2-m_\pi^2}\ 
\left\{
\frac{(4qk_2-2qk_1+2k_1k_2-q^2)}{(q-k_1-k_2)^2-m_\pi^2}
+\frac{3m_\pi^2}{(q-k_1-k_2)^2-m_\pi^2}\right\}
2M\,\overline{u}(\vec{p}\,')
\gamma_5\,
u(\vec{p})
\end{equation}
\begin{equation}
{\cal A}^\mu_e=-\frac{1}{f_\pi^2}\cos\theta_c\ \frac{2k_1^\mu-q^\mu}{(q-k_1)^2-m_\pi^2}\
2F_1^V\,\overline{u}(\vec{p}\,')
\slashchar{k}_2\
\,u(\vec{p})
\end{equation}
\begin{equation}
{\cal A}^\mu_{a'} = \frac{1}{4f_\pi^2}\cos\theta_c\ 
\overline{u}(\vec{p}\,')\left(
\slashchar{k}_1-\slashchar{k}_2\right) S(p+q)
\left(V^\mu - A^\mu\right)
u(\vec{p})
\end{equation}
\begin{equation}
{\cal A}^\mu_{b'} = -\frac{1}{4f_\pi^2}\cos\theta_c\ 
\overline{u}(\vec{p}\,')\left(V^\mu - A^\mu \right) 
S(p'-q)\left(\slashchar{k}_1-\slashchar{k}_2\right)
u(\vec{p})
\end{equation}
\begin{equation}
{\cal A}^\mu_f=\frac{g_A}{2 f_\pi^2}\cos\theta_c\ 
\overline{u}(\vec{p}\,')\left(
\slashchar{k}_2\gamma_5S(p'+k_2)
2F_1^Vg_A\gamma^\mu\gamma_5-
\slashchar{k}_2\gamma_5S(p'+k_2)\gamma^\mu\right)u(\vec{p})
\end{equation}
\begin{equation}
{\cal A}^\mu_g=-\frac{g_A}{2 f_\pi^2}\cos\theta_c\ 
\overline{u}(\vec{p}\,')\left( 2F_1^Vg_A\gamma^\mu\gamma_5S(p-k_2)\slashchar{k}_2\gamma_5\
-\gamma^\mu
S(p-k_2)\slashchar{k}_2\gamma_5\,\right)
u(\vec{p})
\end{equation}
\begin{equation}
{\cal A}^\mu_h=\frac{g_A}{4f_\pi^2}\cos\theta_c\ \frac{q^\mu}{q^2-m_\pi^2}\
\overline{u}(\vec{p}\,') \slashchar{k}_2\gamma_5 S(p'+k_2)
(\slashchar{q}+\slashchar{k}_1)u(\vec{p})
\end{equation}
\begin{equation}
{\cal A}^\mu_i=-\frac{g_A}{4f_\pi^2}\cos\theta_c\ \frac{q^\mu}{q^2-m_\pi^2}\
\overline{u}(\vec{p}\,')
(\slashchar{q}+\slashchar{k}_1)S(p-k_2)
\slashchar{k}_2\gamma_5\,
u(\vec{p})
\end{equation}
\begin{equation}
{\cal A}^\mu_j=-\frac{g_A^2}{2f_\pi^2}\cos\theta_c\ 2F_1^V\frac{q^\mu-2k_1^\mu}{(q-k_1)^2-m_\pi^2}\
\overline{u}(\vec{p}\,')\slashchar{k}_2\gamma_5
S(p'+k_2)(\slashchar{q}-\slashchar{k}_1)\gamma_5\,
u(\vec{p})
\end{equation}
\begin{equation}
{\cal A}^\mu_k=\frac{g_A^2}{2f_\pi^2}\cos\theta_c\ 2F_1^V
\frac{q^\mu-2k_1^\mu}{(q-k_1)^2-m_\pi^2}\
\overline{u}(\vec{p}\,')(\slashchar{q}-\slashchar{k}_1)\gamma_5
S(p-k_2)\slashchar{k}_2\gamma_5\,u(\vec{p})
\end{equation}
\begin{equation}
{\cal A}^\mu_l=-\frac{g_A^2}{2f_\pi^2}\cos\theta_c\ 
\overline{u}(\vec{p}\,') \slashchar{k}_2\gamma_5
S(p'+k_2)\slashchar{k}_1\gamma_5S(p'+k_1+k_2)
\left( V^\mu - A^\mu \right)
u(\vec{p})
\end{equation}
\begin{equation}
{\cal A}^\mu_n=-\frac{g_A^2}{2f_\pi^2}\cos\theta_c\ 
\overline{u}(\vec{p}\,')
\left( V^\mu - A^\mu \right)
S(p-k_1-k_2) \slashchar{k}_1\gamma_5
S(p-k_2)\slashchar{k}_2\gamma_5\,
u(\vec{p})
\end{equation}
\subsection{$W^+_qn_p\to n_{p'}\pi^+_{k_1}\pi^0_{k_2}$}
\begin{equation}
{\cal A}^\mu_a=-\frac{1}{2\sqrt2f_\pi^2}\cos\theta_c\ \overline{u}(\vec{p}\,')
\left(2F_1^V\,\gamma^\mu-g_A\,\gamma^\mu\gamma_5
\right)
u(\vec{p})
\end{equation}
\begin{equation}
{\cal A}^\mu_b=-\frac{g_A}{6\sqrt2f_\pi^2}\cos\theta_c\ \frac{q^\mu}{q^2-m_\pi^2}\,\
\overline{u}(\vec{p}\,')
\left(3\,\slashchar{k}_2+2M\,\right)\gamma_5\,
u(\vec{p})
\end{equation}
\begin{equation}
{\cal A}^\mu_c=-\frac{g_A\sqrt2}{3f_\pi^2}\cos\theta_c\ \frac{q^\mu-3k_1^\mu}{(q-k_1-k_2)^2-m_\pi^2}\
2M\,\overline{u}(\vec{p}\,')
\gamma_5\,
u(\vec{p})
\end{equation}
\begin{equation}
{\cal A}^\mu_d=-\frac{g_A}{3\sqrt2f_\pi^2}\cos\theta_c\ \frac{q^\mu}{q^2-m_\pi^2}\
\frac{4qk_1-2qk_2+2k_1k_2-q^2}{(q-k_1-k_2)^2-m_\pi^2}\,
2M\,\overline{u}(\vec{p}\,')
\gamma_5\,
u(\vec{p})
\end{equation}
\begin{equation}
{\cal A}^\mu_e=-\frac{1}{\sqrt2f_\pi^2}\cos\theta_c\ \frac{2k_2^\mu-q^\mu}{(q-k_2)^2-m_\pi^2}\
2F_1^V\,\overline{u}(\vec{p}\,')
\slashchar{k}_1\
\,u(\vec{p})
\end{equation}
\begin{equation}
{\cal A}^\mu_{a'} = -\frac{\sqrt2}{4f_\pi^2}\cos\theta_c\ 
\overline{u}(\vec{p}\,')\left(
\slashchar{k}_1-\slashchar{k}_2\right) S(p+q)
\left(V^\mu - A^\mu\right) u(\vec{p})
\end{equation}
\begin{eqnarray}
{\cal A}^\mu_f=&&-\frac{g_A}{2\sqrt2 f_\pi^2}\cos\theta_c\ \left\{\ \ 
\overline{u}(\vec{p}\,')\left(
\slashchar{k}_2\gamma_5S(p'+k_2)-2\slashchar{k}_1\gamma_5S(p'+k_1)\right)
2F_1^Vg_A\gamma^\mu\gamma_5\,u(\vec{p})\right.\nonumber\\
&&\left.-\overline{u}(\vec{p}\,')\left(
\slashchar{k}_2\gamma_5S(p'+k_2)-2 \slashchar{k}_1\gamma_5S(p'+k_1)
\right)\gamma^\mu\,u(\vec{p})\right\}
\end{eqnarray}
\begin{equation}
{\cal A}^\mu_g=-\frac{g_A}{2\sqrt2 f_\pi^2}\cos\theta_c\ 
\overline{u}(\vec{p}\,')\left(2F_1^Vg_A\gamma^\mu\gamma_5S(p-k_2)\slashchar{k}_2\gamma_5\
-\gamma^\mu S(p-k_2)\slashchar{k}_2\gamma_5\,\right)
u(\vec{p})
\end{equation}
\begin{equation}
{\cal A}^\mu_h=-\frac{g_A}{4\sqrt2f_\pi^2}\cos\theta_c\frac{q^\mu}{q^2-m_\pi^2}
\overline{u}(\vec{p}\,')\left(\slashchar{k}_2\gamma_5 S(p'+k_2)
(\slashchar{q}+\slashchar{k}_1)
-2 \slashchar{k}_1\gamma_5 S(p'+k_1)
(\slashchar{q}+\slashchar{k}_2)\right)
u(\vec{p})
\end{equation}
\begin{equation}
{\cal A}^\mu_i=-\frac{g_A}{4\sqrt2f_\pi^2}\cos\theta_c\ \frac{q^\mu}{q^2-m_\pi^2}\
\overline{u}(\vec{p}\,')
(\slashchar{q}+\slashchar{k}_1)S(p-k_2)
\slashchar{k}_2\gamma_5\, u(\vec{p})
\end{equation}
\begin{eqnarray}
{\cal A}^\mu_j=\frac{g_A^2}{2\sqrt2f_\pi^2}\cos\theta_c\ 2F_1^V\bigg(&&
 \frac{q^\mu-2k_1^\mu}{(q-k_1)^2-m_\pi^2}\
\overline{u}(\vec{p}\,')\slashchar{k}_2\gamma_5
S(p'+k_2)(\slashchar{q}-\slashchar{k}_1)\gamma_5\,
u(\vec{p})\nonumber\\
&&-2\frac{q^\mu-2k_2^\mu}{(q-k_2)^2-m_\pi^2}\
\overline{u}(\vec{p}\,')\slashchar{k}_1\gamma_5
S(p'+k_1)(\slashchar{q}-\slashchar{k}_2)\gamma_5\,
u(\vec{p})
\bigg)
\end{eqnarray}
\begin{equation}
{\cal A}^\mu_k=\frac{g_A^2}{2\sqrt2f_\pi^2}\cos\theta_c\ 
2F_1^V\frac{q^\mu-2k_1^\mu}{(q-k_1)^2-m_\pi^2}\
\overline{u}(\vec{p}\,')(\slashchar{q}-\slashchar{k}_1)\gamma_5
S(p-k_2)\slashchar{k}_2\gamma_5\,
u(\vec{p})
\end{equation}
\begin{equation}
{\cal A}^\mu_l=\frac{g_A^2}{2\sqrt2f_\pi^2}\cos\theta_c\ 
\overline{u}(\vec{p}\,') \left(\slashchar{k}_2\gamma_5S(p'+k_2)
\slashchar{k}_1\gamma_5-\slashchar{k}_1\gamma_5S(p'+k_1)
\slashchar{k}_2\gamma_5\right)S(p'+k_1+k_2)
\left( V^\mu - A^\mu\right)
u(\vec{p})
\end{equation}
\begin{equation}
{\cal A}^\mu_m = \frac{g_A^2}{2\sqrt2f_\pi^2}\cos\theta_c\ 
\overline{u}(\vec{p}\,')\slashchar{k}_1\gamma_5
S(p'+k_1)\left( V^\mu - A^\mu\right)
S(p-k_2)\slashchar{k}_2\gamma_5\,u(\vec{p})
\end{equation}


\begin{thebibliography}{99}

%QE Teor
%\cite{Nieves:2004wx}
\bibitem{Nieves:2004wx}
  J.~Nieves, J.~E.~Amaro and M.~Valverde,
  %``Inclusive quasi-elastic neutrino reactions: Combined study of the
  %inclusive muon capture in C-12 and the LSND measurement of the reaction
  %C-12(nu/mu,mu-)X near threshold,''
  Phys.\ Rev.\  C {\bf 70} (2004) 055503
  [Erratum-ibid.\  C {\bf 72} (2005) 019902]
  [arXiv:nucl-th/0408005].
  %%CITATION = PHRVA,C70,055503;%%

\bibitem{Nieves:2005rq}
  J.~Nieves, M.~Valverde and M.~J.~Vicente Vacas,
  %``Inclusive nucleon emission induced by quasi-elastic neutrino nucleus
  %interactions,''
  Phys.\ Rev.\  C {\bf 73}, 025504 (2006)
  [arXiv:hep-ph/0511204].
  %%CITATION = PHRVA,C73,025504;%%

\bibitem{Valverde:2006zn}
 M.~Valverde, J.~E.~Amaro and J.~Nieves,
 %``Theoretical uncertainties on quasielastic charged-current neutrino  nucleus
 %cross sections,''
 Phys.\ Lett.\  B {\bf 638}, 325 (2006)
 [arXiv:hep-ph/0604042].
 %%CITATION = PHLTA,B638,325;%%


%\cite{Martinez:2005xe}
\bibitem{Martinez:2005xe}
  M.~C.~Martinez, P.~Lava, N.~Jachowicz, J.~Ryckebusch, K.~Vantournhout and J.~M.~Udias,
  %``Relativistic models for quasi-elastic neutrino scattering,''
  Phys.\ Rev.\  C {\bf 73} (2006) 024607
  [arXiv:nucl-th/0505008].
  %%CITATION = PHRVA,C73,024607;%%
  
%\cite{Meucci:2004ip}
\bibitem{Meucci:2004ip}
  A.~Meucci, C.~Giusti and F.~D.~Pacati,
  %``Neutral-current neutrino nucleus quasielastic scattering,''
  Nucl.\ Phys.\  A {\bf 744} (2004) 307
  [arXiv:nucl-th/0405004].
  %%CITATION = NUPHA,A744,307;%%
  
  %\cite{Caballero:2005sj}
\bibitem{Caballero:2005sj}
  J.~A.~Caballero, J.~E.~Amaro, M.~B.~Barbaro, T.~W.~Donnelly, C.~Maieron and J.~M.~Udias,
  %``Superscaling in charged current neutrino quasielastic scattering in the
  %relativistic impulse approximation,''
  Phys.\ Rev.\ Lett.\  {\bf 95} (2005) 252502
  [arXiv:nucl-th/0504040].
  %%CITATION = PRLTA,95,252502;%%

%\cite{Leitner:2006ww}
\bibitem{Leitner:2006ww}
  T.~Leitner, L.~Alvarez-Ruso and U.~Mosel,
  %``Charged current neutrino nucleus interactions at intermediate energies,''
  Phys.\ Rev.\  C {\bf 73} (2006) 065502
  [arXiv:nucl-th/0601103].
  %%CITATION = PHRVA,C73,065502;%%    

% exp. QE

%\cite{Gran:2006jn}
\bibitem{Gran:2006jn}
  R.~Gran {\it et al.}  [K2K Collaboration],
  %``Measurement of the quasi-elastic axial vector mass in neutrino oxygen
  %interactions,''
  Phys.\ Rev.\  D {\bf 74} (2006) 052002
  [arXiv:hep-ex/0603034].
  %%CITATION = PHRVA,D74,052002;%%
  
%\cite{:2007ru}
\bibitem{:2007ru}
  A.~A.~Aguilar-Arevalo {\it et al.}  [MiniBooNE Collaboration],
  %``Measurement of Muon Neutrino Quasi-Elastic Scattering on Carbon  ,''
  arXiv:0706.0926 [hep-ex].
  %%CITATION = ARXIV:0706.0926;%%

%  exp 1 pion 

%\cite{Radecky:1981fn}
\bibitem{Radecky:1981fn}
  G.~M.~Radecky {\it et al.},
  %``Study Of Single Pion Production By Weak Charged Currents In Low-Energy
  %Neutrino D Interactions,''
  Phys.\ Rev.\  D {\bf 25} (1982) 1161
  [Erratum-ibid.\  D {\bf 26} (1982) 3297].
  %%CITATION = PHRVA,D25,1161;%%
  
%\cite{Kitagaki:1990vs}
\bibitem{Kitagaki:1990vs}
  T.~Kitagaki {\it et al.},
  %``Study Of Neutrino D $\to$ Mu- P P(S) And Neutrino D $\to$ Mu- Delta++
  %(1232) N(S) Using The Bnl 7-Foot Deuterium Filled Bubble Chamber,''
  Phys.\ Rev.\  D {\bf 42} (1990) 1331.
  %%CITATION = PHRVA,D42,1331;%%

%theory 1 pion  
  
%\cite{Adler:1968tw}
\bibitem{Adler:1968tw}
  S.~L.~Adler,
  %``Photoproduction, electroproduction and weak single pion production in the
  %(3,3) resonance region,''
  Annals Phys.\  {\bf 50} (1968) 189.
  %%CITATION = APNYA,50,189;%%

%\cite{Schreiner:1973ka}
\bibitem{Schreiner:1973ka}
  P.~A.~Schreiner and F.~Von Hippel,
  %``Nu p $\to$ mu- delta++ - comparison with theory,''
  Phys.\ Rev.\ Lett.\  {\bf 30} (1973) 339.
  %%CITATION = PRLTA,30,339;%%

%\cite{AlvarezRuso:1997jr}
\bibitem{AlvarezRuso:1997jr}
  L.~Alvarez-Ruso, S.~K.~Singh and M.~J.~Vicente Vacas,
  %``Charged current weak electroproduction of Delta resonance,''
  Phys.\ Rev.\  C {\bf 57} (1998) 2693
  [arXiv:nucl-th/9712058].
  %%CITATION = PHRVA,C57,2693;%%

%\cite{AlvarezRuso:1998hi}
\bibitem{AlvarezRuso:1998hi}
  L.~Alvarez-Ruso, S.~K.~Singh and M.~J.~Vicente Vacas,
  %``nu d --> mu- Delta++ n reaction and axial vector N Delta coupling,''
  Phys.\ Rev.\  C {\bf 59} (1999) 3386
  [arXiv:nucl-th/9804007].
  %%CITATION = PHRVA,C59,3386;%%

%\cite{Paschos:2003qr}
\bibitem{Paschos:2003qr}
  E.~A.~Paschos, J.~Y.~Yu and M.~Sakuda,
  %``Neutrino production of resonances,''
  Phys.\ Rev.\  D {\bf 69} (2004) 014013
  [arXiv:hep-ph/0308130].
  %%CITATION = PHRVA,D69,014013;%%

%\cite{Lalakulich:2005cs}
\bibitem{Lalakulich:2005cs}
  O.~Lalakulich and E.~A.~Paschos,
  %``Resonance production by neutrinos. I: J = 3/2 resonances,''
  Phys.\ Rev.\  D {\bf 71} (2005) 074003
  [arXiv:hep-ph/0501109].
  %%CITATION = PHRVA,D71,074003;%%
  
%\cite{Lalakulich:2006sw}
\bibitem{Lalakulich:2006sw}
  O.~Lalakulich, E.~A.~Paschos and G.~Piranishvili,
  %``Resonance production by neutrinos: The second resonance region,''
  Phys.\ Rev.\  D {\bf 74} (2006) 014009
  [arXiv:hep-ph/0602210].
  %%CITATION = PHRVA,D74,014009;%%
  


% theor 1pion backg

%\cite{Fogli:1979cz}
\bibitem{Fogli:1979cz}
  G.~L.~Fogli and G.~Nardulli,
  %``A New Approach To The Charged Current Induced Weak One Pion Production,''
  Nucl.\ Phys.\  B {\bf 160} (1979) 116.
  %%CITATION = NUPHA,B160,116;%%
  
%\cite{Sato:2003rq}
\bibitem{Sato:2003rq}
  T.~Sato, D.~Uno and T.~S.~H.~Lee,
  %``Dynamical model of weak pion production reactions,''
  Phys.\ Rev.\  C {\bf 67} (2003) 065201
  [arXiv:nucl-th/0303050].
  %%CITATION = PHRVA,C67,065201;%%

%\cite{Hernandez:2007qq}
\bibitem{Hernandez:2007qq}
  E.~Hernandez, J.~Nieves and M.~Valverde,
  %``Weak pion production off the nucleon,''
  Phys.\ Rev.\ D {\bf 76} (2007) 033005
  [arXiv:hep-ph/0701149].
  %%CITATION = HEP-PH/0701149;%%
  
% preliminary 1pion data

\bibitem{nuint07}
C. Mariani, 
Proceedings of Nuint07.

%\cite{Wascko:2006tx}
\bibitem{Wascko:2006tx}
  M.~O.~Wascko  [MiniBooNE Collaboration],
  %``Charged current single pion cross section measurement at MiniBooNE,''
  Nucl.\ Phys.\ Proc.\ Suppl.\  {\bf 159} (2006) 50
  [arXiv:hep-ex/0602050].
  %%CITATION = NUPHZ,159,50;%%  

%
\bibitem{link}
  J.~M.~Link  [MiniBooNE Collaboration], arXiv:0709.3213v1 [hep-ex].

  
% coherent 1 pion  

%\cite{Hasegawa:2005td}
\bibitem{Hasegawa:2005td}
  M.~Hasegawa {\it et al.}  [K2K Collaboration],
  %``Search for coherent charged pion production in neutrino carbon
  %interactions,''
  Phys.\ Rev.\ Lett.\  {\bf 95} (2005) 252301
  [arXiv:hep-ex/0506008].
  %%CITATION = PRLTA,95,252301;%%

%\cite{Wascko:2006ty}
\bibitem{Wascko:2006ty}
  M.~O.~Wascko  [MiniBooNE Collaboration],
  %``Prospects for antineutrino running at MiniBooNE,''
  Nucl.\ Phys.\ Proc.\ Suppl.\  {\bf 159} (2006) 79
  [arXiv:hep-ex/0602051].
  %%CITATION = NUPHZ,159,79;%%

%\cite{AlvarezRuso:2007it}
\bibitem{AlvarezRuso:2007it}
  L.~Alvarez-Ruso, L.~S.~Geng and M.~J.~Vicente Vacas,
  %``Neutral current coherent pion production,''
  arXiv:0707.2172 [nucl-th].
  %%CITATION = ARXIV:0707.2172;%%
  
%\cite{Rein:1980wg}
\bibitem{Rein:1980wg}
  D.~Rein and L.~M.~Sehgal,
  %``Neutrino Excitation Of Baryon Resonances And Single Pion Production,''
  Annals Phys.\  {\bf 133} (1981) 79.
  %%CITATION = APNYA,133,79;%%  
  
% EM form factors resonances
  
%\cite{Burkert:2004sk}
\bibitem{Burkert:2004sk}
  V.~D.~Burkert and T.~S.~H.~Lee,
  %``Electromagnetic meson production in the nucleon resonance region,''
  Int.\ J.\ Mod.\ Phys.\  E {\bf 13} (2004) 1035
  [arXiv:nucl-ex/0407020].
  %%CITATION = IMPAE,E13,1035;%%

%\cite{Aznauryan:2004jd}
\bibitem{Aznauryan:2004jd}
  I.~G.~Aznauryan, V.~D.~Burkert, H.~Egiyan, K.~Joo, R.~Minehart and L.~C.~Smith,
  %``Electroexcitation of the P33(1232), P11(1440), D13(1520), S11(1535) at
  %Q**2 = 0.4-(GeV/c)**2 and 0.65-(GeV/c)**2,''
  Phys.\ Rev.\  C {\bf 71} (2005) 015201
  [arXiv:nucl-th/0407021].
  %%CITATION = PHRVA,C71,015201;%%

%\cite{Tiator:2003uu}
\bibitem{Tiator:2003uu}
  L.~Tiator, D.~Drechsel, S.~Kamalov, M.~M.~Giannini, E.~Santopinto and A.~Vassallo,
  %``Electroproduction of nucleon resonances,''
  Eur.\ Phys.\ J.\  A {\bf 19} (2004) 55
  [arXiv:nucl-th/0310041].
  %%CITATION = EPHJA,A19,55;%%
  

 
%% pi,2pi need for Roper  at threshold
  
%\cite{Oset:1985wt}
\bibitem{Oset:1985wt}
  E.~Oset and M.~J.~Vicente-Vacas,
  %``A Model For The Pi- P $\to$ Pi+ Pi- N Reaction,''
  Nucl.\ Phys.\  A {\bf 446} (1985) 584.
  %%CITATION = NUPHA,A446,584;%%

%idem   
%\cite{Kamano:2004es}
\bibitem{Kamano:2004es}
  H.~Kamano, M.~Morishita and M.~Arima,
  %``Chiral symmetry and N*(1440) --> N pi pi decay,''
  Phys.\ Rev.\  C {\bf 71} (2005) 045201
  [arXiv:nucl-th/0412032].
  %%CITATION = PHRVA,C71,045201;%%

% Roper pRoperties
%\cite{Hernandez:2002xk}
\bibitem{Hernandez:2002xk}
  E.~Hernandez, E.~Oset and M.~J.~Vicente Vacas,
  %``The two pion decay of the Roper resonance,''
  Phys.\ Rev.\  C {\bf 66} (2002) 065201
  [arXiv:nucl-th/0209009].
  %%CITATION = PHRVA,C66,065201;%%
  
  
    
%\cite{Alvarez-Ruso:1997mx}
\bibitem{Alvarez-Ruso:1997mx}
  L.~Alvarez-Ruso, E.~Oset and E.~Hernandez,
  %``Theoretical study of the N N --> N N pi pi reaction,''
  Nucl.\ Phys.\  A {\bf 633} (1998) 519
  [arXiv:nucl-th/9706046].
  %%CITATION = NUPHA,A633,519;%%

%\cite{Alvarez-Ruso:1998xg}
\bibitem{Alvarez-Ruso:1998xg}
  L.~Alvarez-Ruso,
  %``The role of the Roper resonance in n p --> d (pi pi)0,''
  Phys.\ Lett.\  B {\bf 452} (1999) 207
  [arXiv:nucl-th/9811058].
  %%CITATION = PHLTA,B452,207;%%%study of Roper    

%%NEW REFERENCES (REFEREE SUGGESTED)  
%\cite{Allasia:1990uy}
\bibitem{Allasia:1990uy}
  D.~Allasia {\it et al.},
  %``Investigation of exclusive channels in neutrino / anti-neutrino deuteron
  %charged current interactions,''
  Nucl.\ Phys.\  B {\bf 343}, 285 (1990).
  %%CITATION = NUPHA,B343,285;%%

%\cite{Jones:1991tm}
\bibitem{Jones:1991tm}
  G.~T.~Jones {\it et al.},
  %``Inclusive rho0 (770) meson production in neutrino p and anti-neutrino p
  %charged current interactions,''
  Z.\ Phys.\  C {\bf 51} (1991) 11.
  %%CITATION = ZEPYA,C51,11;%%
  
 \bibitem{Wittek:1989tu}
  W.~Wittek {\it et al.}  [BEBC WA59 Collaboration],
  %``PRODUCTION OF rho+, rho-, rho0 (770), eta (550), omega (783) AND f2 (1270)
  %MESONS IN anti-nucleon NEON AND neutrino NEON CHARGED CURRENT INTERACTIONS,''
  Z.\ Phys.\  C {\bf 44}, 175 (1989).
  %%CITATION = ZEPYA,C44,175;%% 
  

%\cite{Grassler:1985cd}
\bibitem{Grassler:1985cd}
  H.~Grassler {\it et al.}  
[Aachen-Birmingham-Bonn-CERN-London-Munich-Oxford
                  Collaboration],
  %``Inclusive Rho0 Production In Anti-Muon-Neutrino P Charged Current
  %Interactions,''
  Nucl.\ Phys.\  B {\bf 272} (1986) 253.
  %%CITATION = NUPHA,B272,253;%%

%\cite{Berge:1979td}
%\bibitem{Berge:1979td}
%  J.~P.~Berge {\it et al.},
%  %``Inclusive Production Of Nonstrange Resonances In High-Energy Neutrino P
%  %Charged Current Interactions,''
%  Phys.\ Rev.\  D {\bf 22} (1980) 1043.
%  %%CITATION = PHRVA,D22,1043

% Exp. low energies

%\cite{Barish:1978pj}
\bibitem{Barish:1978pj}
  S.~J.~Barish {\it et al.},
  %``Study Of Neutrino Interactions In Hydrogen And Deuterium: Inelastic Charged
  %Current Reactions,''
  Phys.\ Rev.\  D {\bf 19} (1979) 2521.
  %%CITATION = PHRVA,D19,2521;%%
  

 

%\cite{Day:1984nf}
\bibitem{Day:1984nf}
  D.~Day {\it et al.},
  %``Study Of Neutrino D Charged Current Two Pion Production In The Threshold
  %Region,''
  Phys.\ Rev.\  D {\bf 28} (1983) 2714.
  %%CITATION = PHRVA,D28,2714;%%

%\cite{Kitagaki:1986ct}
\bibitem{Kitagaki:1986ct}
  T.~Kitagaki {\it et al.},
  %``CHARGED CURRENT EXCLUSIVE PION PRODUCTION IN NEUTRINO DEUTERIUM
  %INTERACTIONS,''
  Phys.\ Rev.\  D {\bf 34} (1986) 2554.
  %%CITATION = PHRVA,D34,2554;%%


% theory

%
%\cite{Biswas:1978ey}
\bibitem{Biswas:1978ey}
  S.~N.~Biswas, S.~R.~Choudhury, A.~K.~Goyal and J.~N.~Passi,
  %``Neutral Current Cross-Sections For Double Pion Production By Neutrinos,''
  Phys.\ Rev.\  D {\bf 18} (1978) 3187.
  %%CITATION = PHRVA,D18,3187;%%


%NC
%\cite{Adjei:1980nj}
\bibitem{Adjei:1980nj}
  S.~A.~Adjei, D.~A.~Dicus and V.~L.~Teplitz,
  %``Neutral Current Two Pion Production From Nucleons,''
  Phys.\ Rev.\  D {\bf 23} (1981) 672.
  %%CITATION = PHRVA,D23,672;%%

%CC 
%\cite{Adjei:1981nw}
\bibitem{Adjei:1981nw}
  S.~A.~Adjei, D.~A.~Dicus and V.~L.~Teplitz,
  %``Charged Current Two Pion Production From Nucleons,''
  Phys.\ Rev.\  D {\bf 24} (1981) 623.
  %%CITATION = PHRVA,D24,623;%%


%\cite{Gomez Tejedor:1993bq}
\bibitem{Gomez Tejedor:1993bq}
  J.~A.~Gomez Tejedor and E.~Oset,
  %``A Model For The Gamma P $\to$ Pi+ Pi- P Reaction,''
  Nucl.\ Phys.\  A {\bf 571} (1994) 667.
  %%CITATION = NUPHA,A571,667;%%  
  
%\cite{Bernard:1995gx}
\bibitem{Bernard:1995gx}
  V.~Bernard, N.~Kaiser and U.~G.~Meissner,
  %``The Reaction Pi N $\to$ Pi Pi N At Threshold In Chiral Perturbation
  %Theory,''
  Nucl.\ Phys.\  B {\bf 457} (1995) 147
  [arXiv:hep-ph/9507418].
  %%CITATION = NUPHA,B457,147;%%



%\cite{Eidelman:2004wy}
\bibitem{Eidelman:2004wy}
  S.~Eidelman {\it et al.}  [Particle Data Group],
  %``Review of particle physics,''
  Phys.\ Lett.\  B {\bf 592} (2004) 1.
  %%CITATION = PHLTA,B592,1;%%

%\cite{Drechsel:2007if}
\bibitem{Drechsel:2007if}
  D.~Drechsel, S.~S.~Kamalov and L.~Tiator,
  %``Unitary Isobar Model - MAID2007,''
  Eur.\ Phys.\ J.\  A {\bf 34}, 69 (2007).
  [arXiv:0710.0306 []].
  %%CITATION = EPHJA,A34,69;%%


  
%\cite{Warns:1989ie}
\bibitem{Warns:1989ie}
  M.~Warns, H.~Schroder, W.~Pfeil and H.~Rollnik,
  %``Calculations Of Electromagnetic Nucleon Form-Factors And Electroexcitation
  %Amplitudes Of Isobars,''
  Z.\ Phys.\  C {\bf 45}, 627 (1990); ibidem 613 (1990)
  %%CITATION = ZEPYA,C45,627;%%


\bibitem{Meyer:2001js}
  U.~Meyer, E.~Hernandez and A.~J.~Buchmann,
  %``Exchange Currents In Nucleon Electroexcitation,''
  Phys.\ Rev.\  C {\bf 64}, 035203 (2001).
  %%CITATION = PHRVA,C64,035203;%%



%\cite{Copley:1969qn}
\bibitem{Copley:1969qn}
  L.~A.~Copley, G.~Karl and E.~Obryk,
  %``Backward single pion photoproduction and the symmetric quark model,''
  Phys.\ Lett.\  B {\bf 29} (1969) 117.
  %%CITATION = PHLTA,B29,117;%%
  
%\cite{Close:1979bt}
\bibitem{Close:1979bt}
  F.~E.~Close,
  ``An Introduction To Quarks And Partons,''
%\href{http://www.slac.stanford.edu/spires/find/hep/www?irn=634026}{SPIRES entry}
{\it  Academic Press/london 1979, 481p} 
  
%\cite{AlvarezRuso:2003gj}
\bibitem{AlvarezRuso:2003gj}
  L.~Alvarez-Ruso, M.~B.~Barbaro, T.~W.~Donnelly and A.~Molinari,
  %``Nuclear response functions for the N - N*(1440) transition,''
  Nucl.\ Phys.\  A {\bf 724} (2003) 157
  [arXiv:nucl-th/0303027].
  %%CITATION = NUPHA,A724,157;%%


%\cite{Galster:1971kv}
\bibitem{Galster:1971kv}
  S.~Galster, H.~Klein, J.~Moritz, K.~H.~Schmidt, D.~Wegener and J.~Bleckwenn,
  %``Elastic electron - deuteron scattering and the electric neutron form-factor
  %at four momentum transfers 5-fm**-2 < q**2 < 14-fm**-2,''
  Nucl.\ Phys.\  B {\bf 32} (1971) 221.
  %%CITATION = NUPHA,B32,221;%%

%\cite{Ericson:1988gk}
\bibitem{Ericson:1988gk}
  T.~E.~O.~Ericson and W.~Weise,
  %``PIONS AND NUCLEI,''
%\href{http://www.slac.stanford.edu/spires/find/hep/www?irn=1965719}{SPIRES entry}
{\it  Oxford, UK: Clarendon (1988) 479 P. (The international series of monographs on physics, 74)}







\end{thebibliography}
\end{document}